\documentclass[journal=apchd5,manuscript=article]{achemso}
\pdfoutput=1
\usepackage[T1]{fontenc} % Use modern font encodings
\usepackage{graphicx}
\usepackage{dcolumn}
\usepackage[separate-uncertainty = true]{siunitx} 
\usepackage{amsmath}

\graphicspath{{figures/pngs}}

 \title{Predicting the structural colors of films of disordered photonic balls}

 \author{Anna B. Stephenson}
 \author{Ming Xiao}
 \affiliation{Harvard John A. Paulson School of Engineering and Applied Sciences, Harvard University, Cambridge, Massachusetts 02138, USA}
 \altaffiliation{Current address: College of Polymer Science and Engineering, Sichuan University, Chengdu 610065, China}
 \author{Victoria Hwang}
 \affiliation{Harvard John A. Paulson School of Engineering and Applied Sciences, Harvard University, Cambridge, Massachusetts 02138, USA}
 \author{Liangliang Qu}
 \author{Paul A. Odorisio}
 \affiliation{BASF Corporation, Tarrytown, New York 10591, USA}
 \altaffiliation{Deceased January 14, 2022.}
 \author{Michael Burke}
 \author{Keith Task}
 \author{Ted Deisenroth}
 \affiliation{BASF Corporation, Tarrytown, New York 10591, USA}
 \author{Solomon Barkley}
 \affiliation{Department of Physics, Harvard University, Cambridge, Massachusetts 02138, USA}
 \author{Rupa H. Darji}
 \affiliation{BASF Corporation, Tarrytown, New York 10591, USA}
 \author{Vinothan N. Manoharan}
 \email{vnm@seas.harvard.edu}
 \affiliation{Harvard John A. Paulson School of Engineering and Applied
    Sciences, Harvard University, Cambridge,
    Massachusetts 02138, USA} 
 \alsoaffiliation{Department of Physics, Harvard University, Cambridge, Massachusetts 02138, USA}


\begin{document}
 
 \begin{abstract}    
   Photonic balls are spheres tens of micrometers in diameter containing
   assemblies of nanoparticles or nanopores with a spacing comparable to
   the wavelength of light. When these nanoscale features are
   disordered, but still correlated, the photonic balls can show
   structural color with low angle-dependence. Their colors, combined
   with the ability to add them to a liquid formulation, make photonic
   balls a promising new type of pigment particle for paints, coatings,
   and other applications. However, it is challenging to predict the
   color of materials made from photonic balls, because the sphere
   geometry and multiple scattering must be accounted for. To address
   these challenges, we develop a multiscale modeling approach involving
   Monte Carlo simulations of multiple scattering at two different
   scales: we simulate multiple scattering and absorption within a
   photonic ball and then use the results to simulate multiple
   scattering and absorption in a film of photonic balls. After
   validating against experimental spectra, we use the model to show
   that films of photonic balls scatter light in fundamentally different
   ways than do homogeneous films of nanopores or nanoparticles, because
   of their increased surface area and refraction at the interfaces of
   the balls. Both effects tend to sharply reduce color saturation
   relative to a homogeneous nanostructured film. We show that saturated
   colors can be achieved by placing an absorber directly in the
   photonic balls and mitigating surface roughness. With these design
   rules, we show that photonic-ball films have an advantage over
   homogeneous nanostructured films: their colors are even less
   dependent on the angle.
 \end{abstract}

\begin{tocentry}
\includegraphics{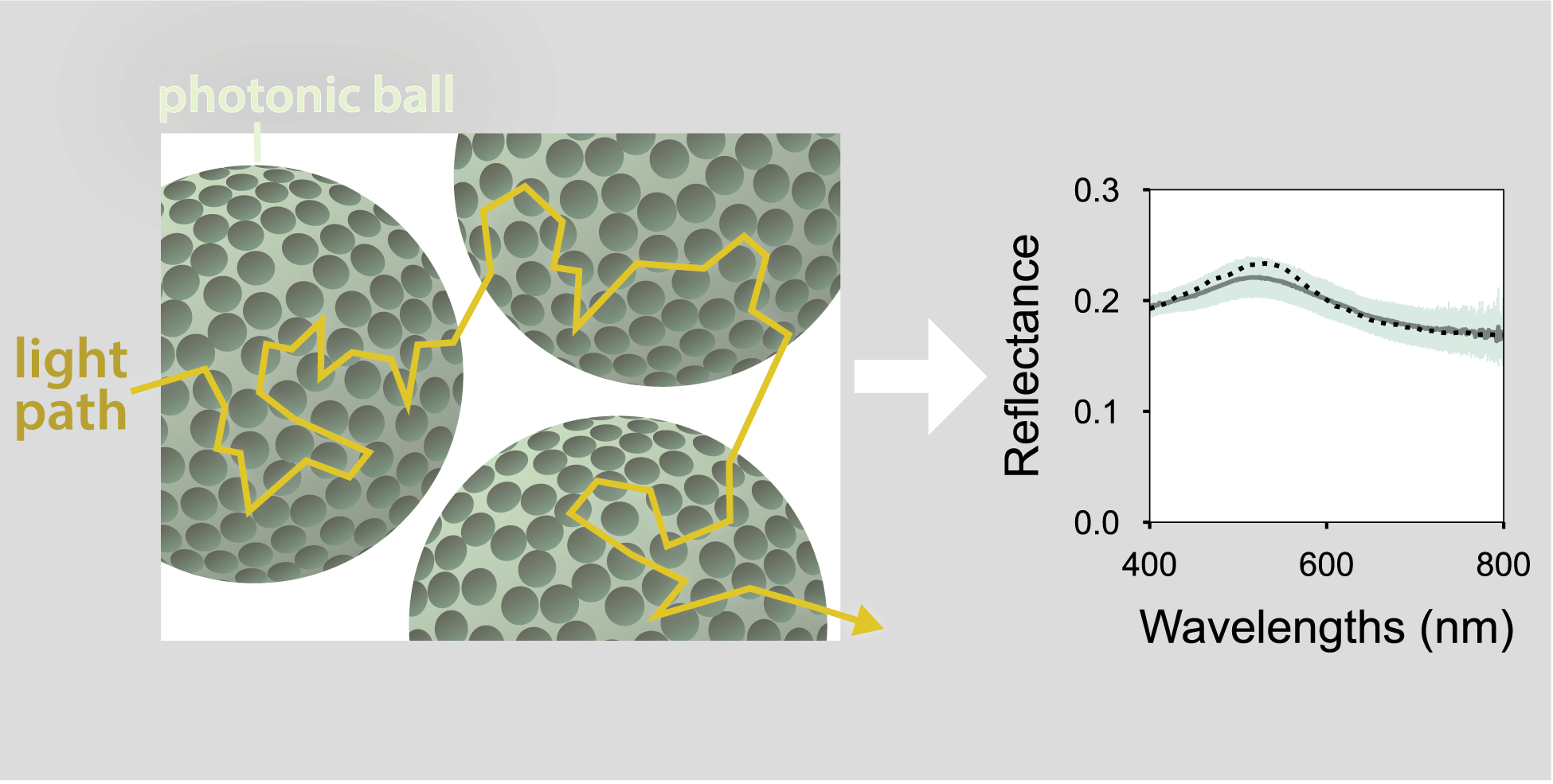} 
For Table of Contents Use Only \\

Manuscript Title: Predicting the structural colors of films of disordered photonic balls \\

Manuscript Authors: Anna B. Stephenson, Ming Xiao, Victoria Hwang, Liangliang Qu, Paul A. Odorisio, Michael Burke, Keith Task, Ted Deisenroth, Solomon Barkley, Rupa H. Darji, Vinothan N. Manoharan \\

Brief Synopsis: Diagram depicting a model for light transport in a
photonic ball. At left is a simulated trajectory of light scattering in
a photonic ball. The results of many simulated trajectories are used to
calculate a reflectance spectrum, shown on the right (dotted line) and
compared to an experimentally measured reflectance spectrum (solid
line).
\end{tocentry}
 
% dissertation copy start here
%%%%%%%%%%%%%%%%%%%%%%%%%%%%%%%%%%%%%%%
\section*{Introduction}
\begin{figure*} 
\includegraphics[width=\textwidth]{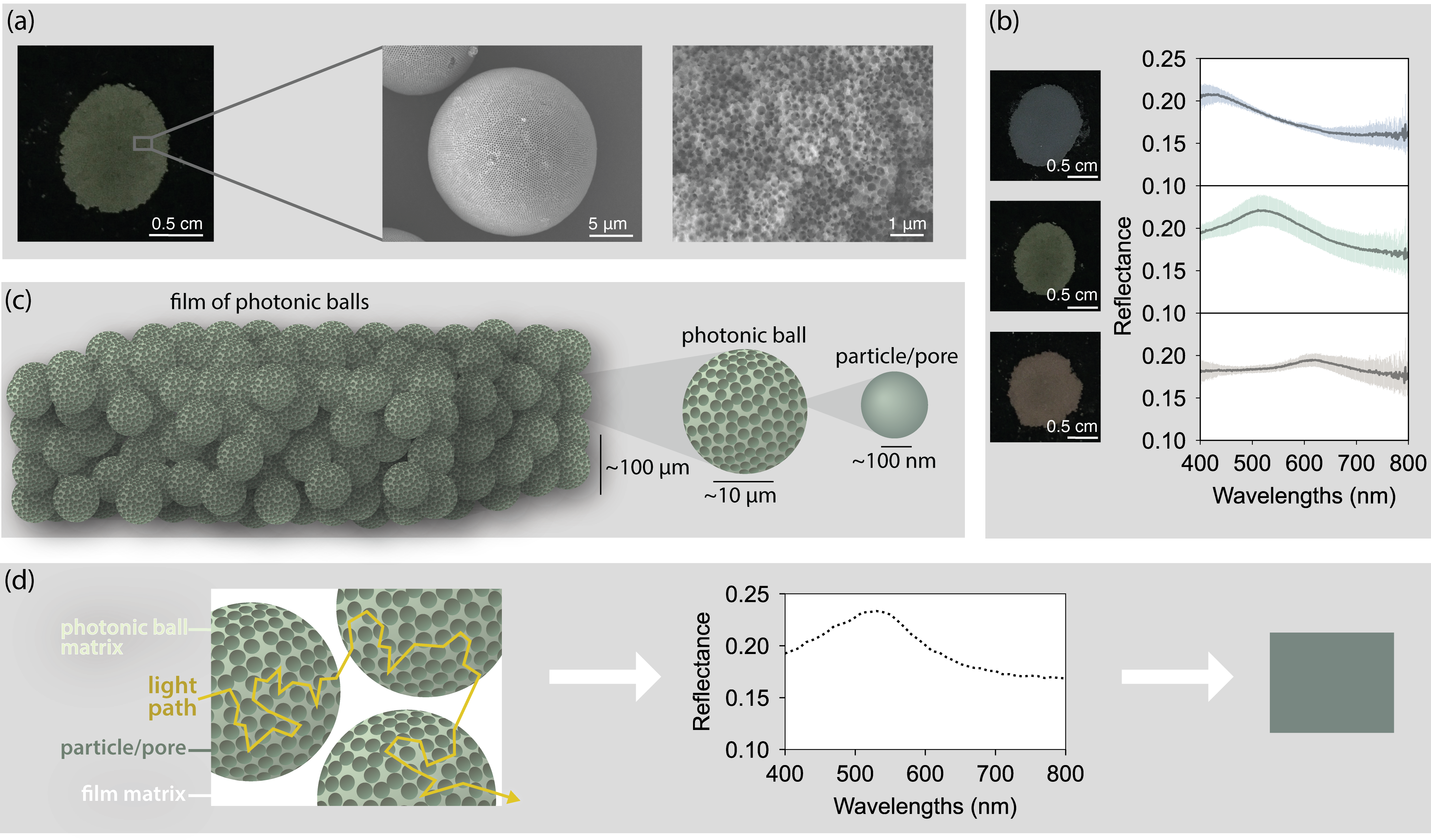} 
\centering
\caption{The geometry of a film of photonic balls leads to complex light
  scattering. (a) Photograph of a green photonic-ball film (left),
  scanning electron micrograph of a representative photonic ball in the
  film (middle), scanning electron micrograph showing the interior of a
  representative photonic ball (right). The photonic balls are made from
  a silica matrix with air nanopores of radius \SI{130}{\nm}. The
  micrographs show that although the outermost layer of the ball is ordered,
  the interior is disordered. (b) Photographs of three photonic-ball
  films (left) accompanied by their reflectance spectra (right). Error
  bars indicate two standard deviations from the mean of 8--11
  measurements across the films. The photonic balls measured are made
  from a silica matrix with nanopores of radius \SI{99}{\nm} (top),
  \SI{130}{\nm} (middle), \SI{163}{\nm} (bottom). (c) Diagram of a film
  of photonic balls showing the relative length scales of the film,
  photonic balls, and nanopores or nanoparticles. (d) At left, diagram
  showing the components of a film of photonic balls and a schematic of
  a path taken by light as it scatters both inside and among the
  photonic balls. We model these paths to calculate the reflectance
  spectrum (middle) and color (right) of the film.}
\label{overview}
\end{figure*}

% introduction word limit: 1000 words
%%%%%%%%%%%%%%%%%%%%%%%%%%%%%%%%%%%%%%%
Photonic balls---also called photonic
microspheres~\cite{rastogi_synthesis_2008},
microcapsules~\cite{kim_osmotic-pressure-controlled_2014},
supraballs~\cite{xiao_bioinspired_2017}, or
supraparticles~\cite{klein_synthesis_2005}---consist of nanoparticles or
nanopores that are packed into micrometer-scale
spheres~\cite{velev_class_2000, moon_electrospray-assisted_2004,
  klein_synthesis_2005, rastogi_synthesis_2008, yu_triphase_2012,
  takeoka_production_2013, zhao_spherical_2014,
  kim_osmotic-pressure-controlled_2014, park_full-spectrum_2014,
  yoshioka_production_2014, teshima_preparation_2015, vogel_color_2015,
  xiao_bioinspired_2017, wang_hollow_2019, ohnuki_grating_2019,
  song_hierarchical_2019, wang_structural_2020, ohnuki_optical_2020,
  sakai_monodisperse_2020, kim_controlled_2020, hu_structurally_2020,
  lim_transparent_2020, zhao_angular-independent_2020,
  yazhgur_light_2021} (Fig.~\ref{overview}a). They are called
``photonic'' because the spacing between the particles or nanopores is
on the order of the wavelength of light, resulting in constructive
interference and structural color (Fig.~\ref{overview}b). Of particular
interest are photonic balls where the nanoscale features are disordered,
but still correlated. These disordered photonic balls have structural
colors that vary only weakly with the sample orientation and angle of
illumination~\cite{takeoka_production_2013,
  kim_osmotic-pressure-controlled_2014, park_full-spectrum_2014,
  yoshioka_production_2014, teshima_preparation_2015,
  xiao_bioinspired_2017, hu_structurally_2020, lim_transparent_2020,
  zhao_angular-independent_2020, yazhgur_light_2021,
  wang_deconvoluting_2022}, because the constructive interference
condition is partially met at a range of wavelengths. Compared to the
more iridescent, or angle-dependent, colors produced by crystalline
photonic balls~\cite{velev_class_2000, moon_electrospray-assisted_2004,
  klein_synthesis_2005, rastogi_synthesis_2008, yu_triphase_2012,
  zhao_spherical_2014, vogel_color_2015, ohnuki_grating_2019,
  song_hierarchical_2019, wang_hollow_2019, ohnuki_optical_2020,
  sakai_monodisperse_2020, kim_controlled_2020, wang_structural_2020},
the colors of disordered photonic balls can be nearly indistinguishable
from those of dyes. Their weak angle-dependence, combined with the
ability to incorporate them in formulations such as liquid suspensions,
powders, or coating precursors, make disordered photonic balls a
promising way to create customizable colors for many different
industrial-scale applications.

The challenge is to determine what color a photonic ball will produce in
a given formulation. The color is determined by the nanoparticle or
nanopore size, the structure of the nanoparticles or nanopores, the
photonic ball size, and the refractive index of the materials that
compose the photonic ball as well as the media in the rest of the
formulation (Fig.~\ref{overview}c). Because of the size of this design
space, it is not possible to fabricate every combination of parameters.
Instead, a predictive model is needed (Fig.~\ref{overview}d).

Some groups have used simulations and modeling to gain a physical
understanding of the scattering from these disordered assemblies of
nanoparticles or nanopores. Single-scattering models, in which light
scatters once from a nanostructured film, are useful for guiding design,
because they are parameterized in terms of the sample
variables~\cite{magkiriadou_absence_2014, maiwald_ewald_2018,
  hwang_effects_2020}, but they do not, in general, quantitatively
reproduce measured spectra because they do not include multiple
scattering, and they do not capture the effects of the photonic-ball
geometry on the color. Numerical methods such as finite-difference
time-domain and finite-integration
time-domain~\cite{dong_structural_2010, shang_highly_2019,
  jacucci_limitations_2020, wang_deconvoluting_2022} can account for
multiple scattering and photonic ball geometry, but these techniques are
computationally intensive and require specifying the positions of every
nanoparticle or nanopore. For systems in which only the volume fraction
and sizes of the nanoscale features are known, an effective-medium
approach is easier to apply and much less computationally intensive.
Schertel and colleagues combined a physically realistic effective-medium
theory with a diffusion approximation~\cite{schertel_structural_2019}.
This model works well for highly multiply scattering systems, but when
designing saturated colors, our principal goal is to reduce multiple
scattering to the point that the diffusion approximation is no longer
valid.

In addition to accounting for weak to moderate multiple scattering, a
model must also account for the effect of the photonic ball's geometry
on its interaction with light. Patil and colleagues have developed
models combining molecular dynamics
simulations~\cite{patil_structural_2022} or computational
reverse-engineering analysis for scattering
experiments~\cite{patil_modeling_2022} with finite-difference
time-domain calculations. Though these models do account for the
photonic ball geometry, we seek a model that is less computationally
intensive and which does not require specifying the positions of all the
nanoscale features. A promising development along these lines is the
single-scattering model developed by Yazhgur and
colleagues~\cite{yazhgur_light_2021} to predict the optical response of
individual photonic balls. Their model has provided a new understanding
of the interplay between the contributions of the nanostructured
scatterers inside the ball and the Mie resonances of the photonic ball
itself. However, this approach does not yet include multiple scattering.
For our purposes, it is essential to account for multiple scattering not
only within a single photonic ball but also within a film of packed
photonic balls, as would be found in many applications.

\begin{figure*}
\centering\includegraphics[width=\textwidth]{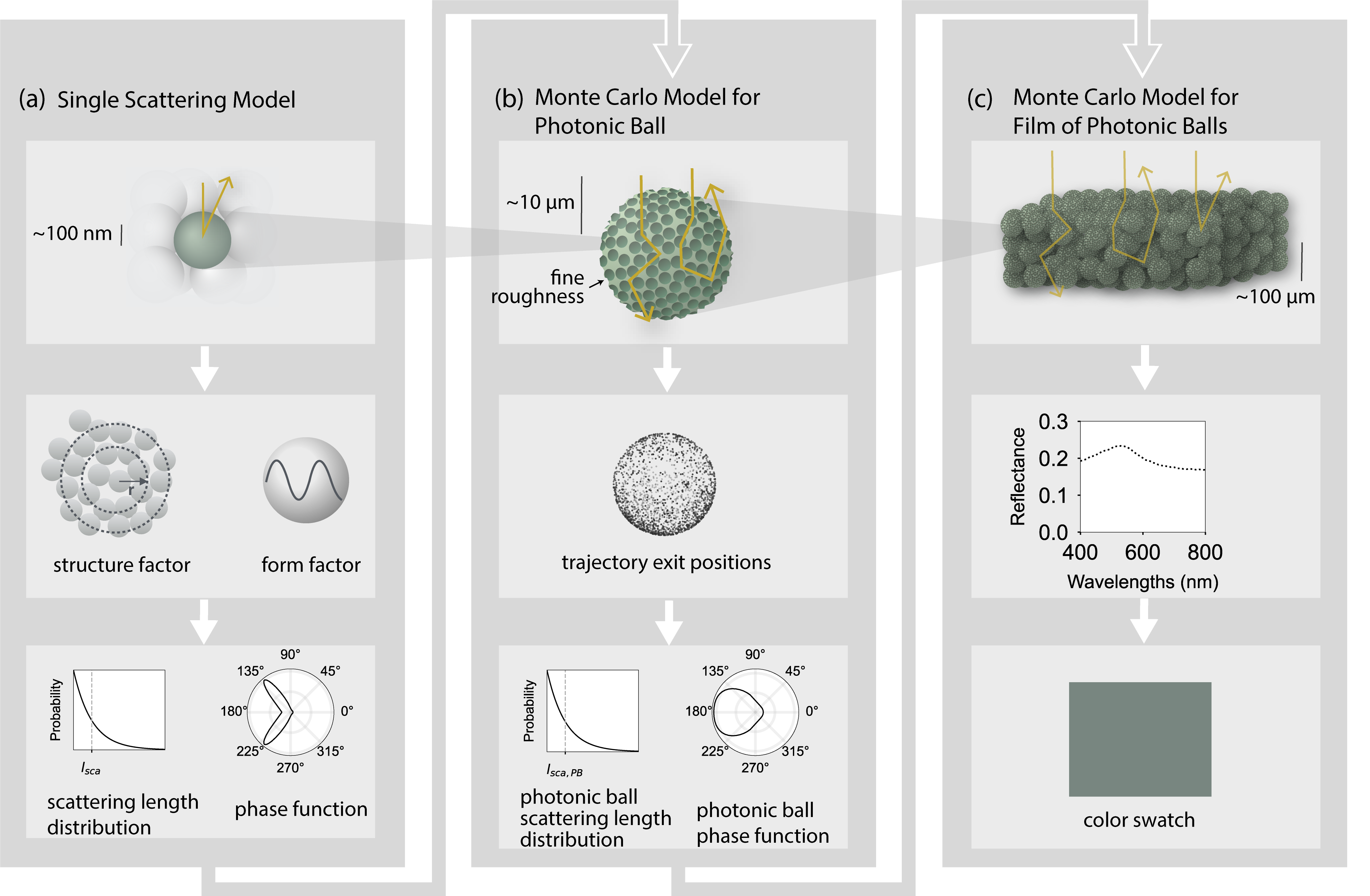}
\caption{Overview of the multiscale model. (a) We first simulate light
scattering from a nanoparticle or nanopore embedded in a disordered
assembly, which we model as an effective medium. We use the nanoparticle
size, refractive index, and effective-medium index to calculate the
structure factor, form factor, and total scattering cross-section. These
quantities determine the scattering-length distribution and phase
function. (b) In the Monte Carlo simulation for a photonic ball, we
sample from these distributions to simulate trajectories of light. We
then calculate a scattering-length distribution and phase function for a
photonic ball using the exit positions of the simulated trajectories.
(c) In the Monte Carlo simulation for a film of photonic balls, we
sample from the photonic-ball distributions to simulate trajectories of
light. We then calculate the reflectance spectrum by summing the
trajectories that exit over a range of angles for each wavelength. The
reflectance spectrum is then converted to a color according to the CIE
standard.
\label{model_overview}}
\end{figure*}

To address these challenges, we develop a multiscale Monte Carlo model
that can be used to predict the color of a composite film of photonic
balls (Fig.~\ref{model_overview}). Our model focuses on a weak multiple
scattering regime---in which the refractive index contrast is high
enough that we must consider multiple scattering but not so high that we
must consider near-field effects. We first use a Monte Carlo calculation
to simulate the light scattering and absorption in an individual
photonic ball. This first calculation is based on previous work by Hwang
and coworkers~\cite{hwang_designing_2021}, though we modify that model
to account for the effects of the photonic-ball geometry. We then use
the results of this simulation in a second Monte Carlo calculation that
simulates the scattering and absorption among multiple photonic balls in
a film. We validate our model against experimental spectra and compare
its predictions to those of other models, showing that the two scales of
simulation and boundary conditions are necessary to achieve experimental
agreement. Our calculations show that the scattering in a photonic-ball
film has fundamental differences from scattering in a nanostructured
film, and those differences lead to different optical effects. We
explore the physical origins of these differences and show that taking
them into consideration brings in new design parameters: the packing
fraction of the photonic balls, the material between the photonic balls,
the thickness of the film, and, importantly, the location of any
absorbing material---in the nanoparticles, photonic-ball matrix, or film
matrix. By exploring the effects of these new design parameters, we
develop design rules for fabricating structurally colored materials for
applications.

%%%%%%%%%%%%%%%%%%
\section*{Results and Discussion}
%%%%%%%%%%%%%%%%%%

\subsection*{Overview of multiscale modeling approach}
In brief, our multiscale model involves two coupled Monte Carlo
simulations (Fig.~\ref{model_overview}). The first simulates light
transport (scattering and absorption) within a photonic ball, and the
second simulates transport in a film of many photonic balls. To
model transport in an individual photonic ball, we simulate light
trajectories by sampling probability distributions for the scattering
length and direction. These distributions are calculated from Mie theory
and effective-medium theory, and they also include the effects of
constructive interference due to structural correlations. To model light
transport in a film of photonic balls, we use the same basic approach,
but the probability distributions are calculated from the results of the
simulation for a single photonic ball. The physical rules taken into
account at each scale are the key to accurately modeling these
materials. We discuss the modeling at both scales in detail in the
Methods section.

%%%%%%
\subsection*{Model validation}
%%%%%%

\begin{figure} 
\centering\includegraphics{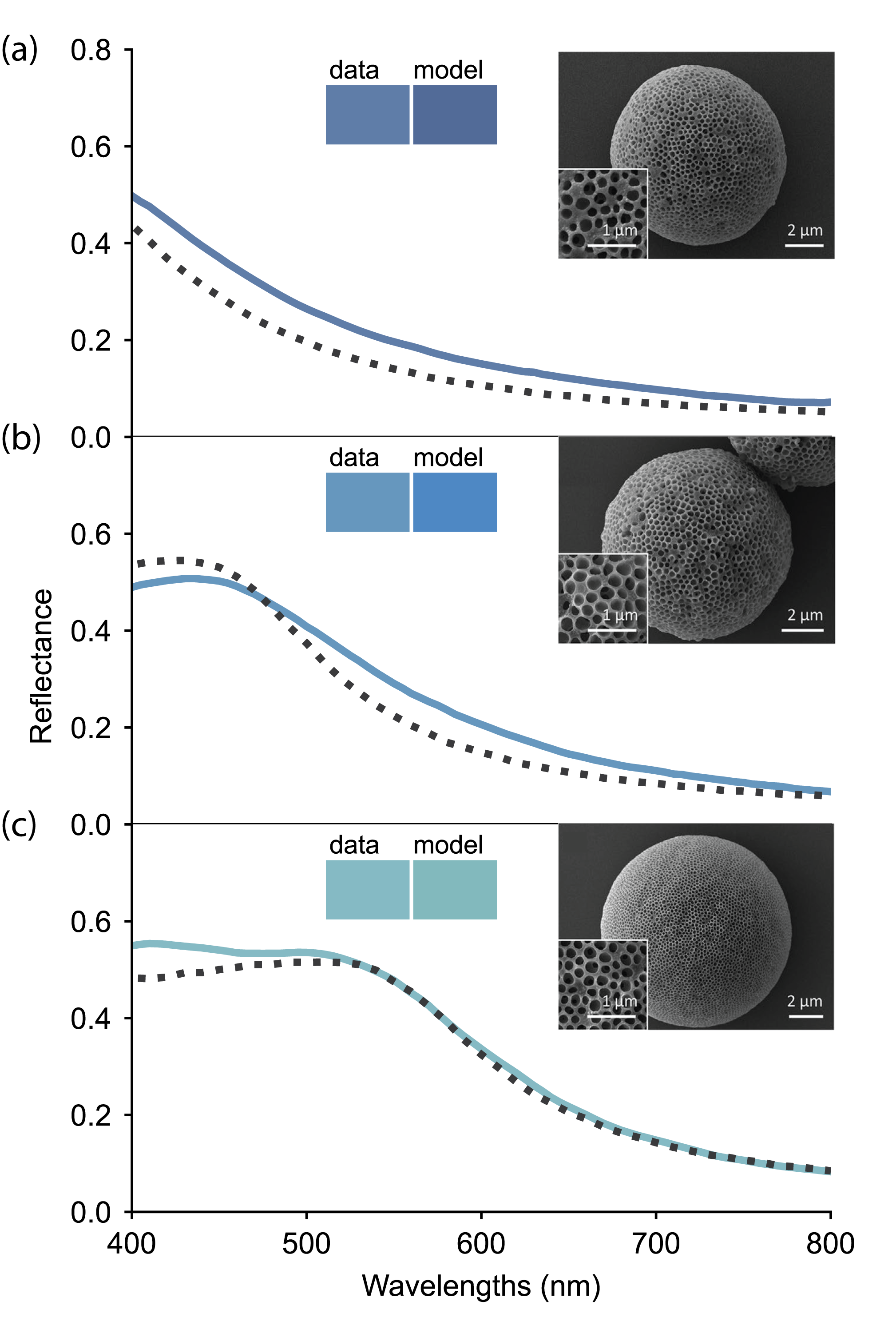}
\caption{Measurements validate the Monte Carlo model for individual
  photonic balls. Measured (solid lines) and predicted (dotted lines)
  reflectance spectra for photonic balls with a primary reflectance peak
  in the (a) ultraviolet, (b) blue, and (c) green. Measurements are from
  Ref.~\citenum{zhao_angular-independent_2020}, Figure S3 (dry samples).
  Left insets show colors calculated from the experimental and predicted
  spectra using the CIELAB colorspace. Right insets show scanning
  electron micrographs of a representative photonic ball for each color,
  adapted with permission from
  Figure~2 of Ref.~\citenum{zhao_angular-independent_2020}, which is
  published under a CC-BY 4.0 license
  (\url{https://creativecommons.org/licenses/by/4.0/}). The model
  parameters are: (a) nanopore radius
  \SI{78}{\nm}, photonic ball diameter \SI{40}{\um}, (b) nanopore radius
  \SI{103}{\nm}, photonic ball diameter \SI{19.9}{\um}, and (c) nanopore
  radius \SI{123}{\nm}, photonic ball diameter \SI{19.3}{\um}. All three
  simulations use a nanopore volume fraction of 0.5, a fine roughness of
  0.01, a matrix refractive index of 1.52, and a nanopore and medium
  refractive index of 1.
  \label{validation_pb}}
\end{figure}

To validate our multiscale Monte Carlo model, we compare the predictions
of our model to three sets of samples. First, we compare the predictions
of the first level of the model---the Monte Carlo simulation for
individual photonic balls---to experimental reflectance data for
individual photonic balls fabricated by Zhao and
colleagues\cite{zhao_angular-independent_2020}. Next, we compare
predictions from the second level---the Monte Carlo simulation for films
of photonic balls---to experimental data from photonic-ball films that
we fabricate and measure ourselves. Finally, for additional validation,
we compare our predictions to measurements of a set of photonic-ball
films made by Takeoka and colleagues~\cite{takeoka_production_2013}
using a different fabrication technique.

Our validations focus on samples in an \textit{intermediate} range of
refractive-index contrast, which we define as $|
n_{\textrm{particle/pore}}-n_{\textrm{matrix}}|$. In this regime, the
index contrast is high enough for multiple scattering to be important,
but not so high that we must model near-field effects. We discuss this
refractive-index regime and our choice to use effective-medium theory in
more detail in the Supporting information. There we compare experimental
data on scattering strength and reflectance spectra to predictions from
various methods (Figs.~S3 and~S4). Based on these calculations, we
expect our model to yield reasonable predictions for samples with an
index contrast of about 0.5, which includes the systems we examine in
this paper (contrasts between $0.45$ and $0.52$). We do not expect it to
accurately predict the reflectance spectra for samples with lower
contrast, such as polystyrene in water (about $0.26$). In the
low-contrast regime, it may be better to avoid an effective-medium
description or to use models intended for the single-scattering regime,
such as models developed by Yazhgur and
colleagues~\cite{yazhgur_light_2021, yazhgur_scattering_2022}.

We validate the first level of the multiscale Monte Carlo model by
comparing the predicted reflectance of individual photonic balls to
experimental measurements from Zhao and
colleagues~\cite{zhao_angular-independent_2020}, who fabricated and
characterized disordered photonic balls and also performed precisely
normalized measurements of the reflectance of individual photonic balls.
We use the data from this study because precise, quantitative
reflectance measurements of a single photonic ball require careful
configuration of the microscope's optical setup to capture only the
reflected light from the photonic ball in view, and the measurements
must be normalized under exactly the same conditions. The measurements
by Zhao and colleagues meet these criteria. The photonic balls used in
this study are inverse structures, with nanopores embedded in a matrix
made from an amphiphilic bottlebrush block copolymer.

To simulate the reflectance spectrum, we must estimate some model
parameters by comparing the model predictions to the data. We vary the
nanopore radius and photonic-ball diameter to lie within two standard
deviations (based on the reported experimental uncertainties) of the
measured values in Ref.~\citenum{zhao_angular-independent_2020}. Because
we do not have experimentally measured values for the volume fraction,
fine roughness, and detection angle range, we choose prior ranges for
these parameters based on physical or experimental considerations and
then estimate the parameters by comparing model predictions to
experiment (see Supporting Information for full details on parameter
estimation). We do not, however, vary the refractive indices. For the
matrix refractive index, we use Zhao \textit{et al.}'s estimated value
of 1.52, which they calculated from reported values from constituent
side chains of the block copolymer. We take the refractive index of the
nanopores to be 1, since the photonic balls are dry.

We find that for each color, the predicted and measured peak locations
line up well (Fig.~\ref{validation_pb}). The deviations between model
and experiment might be due to uncertainty in the photonic-ball size or
other estimated parameters, or they might arise because we do not
include dispersion of the refractive index of the bottlebrush block
copolymer (which, to the best of our knowledge, has not been measured).
Because the refractive index we use is independent of wavelength, and
spectral features appear at different wavelengths for each nanopore
size, we expect the mismatch between data and prediction to vary among
the three samples. But because the model produces good agreement for the
overall magnitude of the reflectance as well as the peak positions for
all three colors, the predicted colors are still close to the measured
colors (see swatches in Fig.~\ref{validation_pb}). Though our model
includes adjustable parameters, these parameters all come from physical
characteristics of the system and are restricted to reasonable physical
ranges. These results show that it is possible to reproduce the main
reflectance features for a single photonic ball, which allows us to
extend this model to a film of photonic balls.

Next we compare the predictions from our model to experimental
measurements of films of photonic balls, validating the second level of
our multiscale model. Reflectance measurements for a film are easier to
perform than measurements on an individual photonic ball, and therefore
we do these measurements in our own setup, using photonic balls that we
fabricate and pack into films (see Methods). These photonic balls
consist of nanopores inside a silica matrix. We fabricate photonic balls
with reflectance peaks in the blue, green, and red
(Fig.~\ref{overview}b) by changing the nanopore size. We then add carbon
black to the photonic balls to suppress multiple scattering and increase
saturation, using a method that likely deposits most of the carbon black
on the photonic-ball surface (see Methods). We measure the carbon-black
concentration, nanopore size, nanopore polydispersity, photonic ball
size, and film thickness to use as input parameters for our model. For
the refractive indices of the silica and carbon black, we use
measurements from the literature~\cite{malitson_interspecimen_1965,
  dalmeida_atmospheric_1991, chylek_effect_1995, kitamura_optical_2007}.
We choose the nanopore volume fraction, photonic-ball volume fraction,
fine roughness, and coarse roughness to lie within reasonable physical
ranges for our samples (see Supporting
Information). 

\begin{figure*} 
  \centering\includegraphics{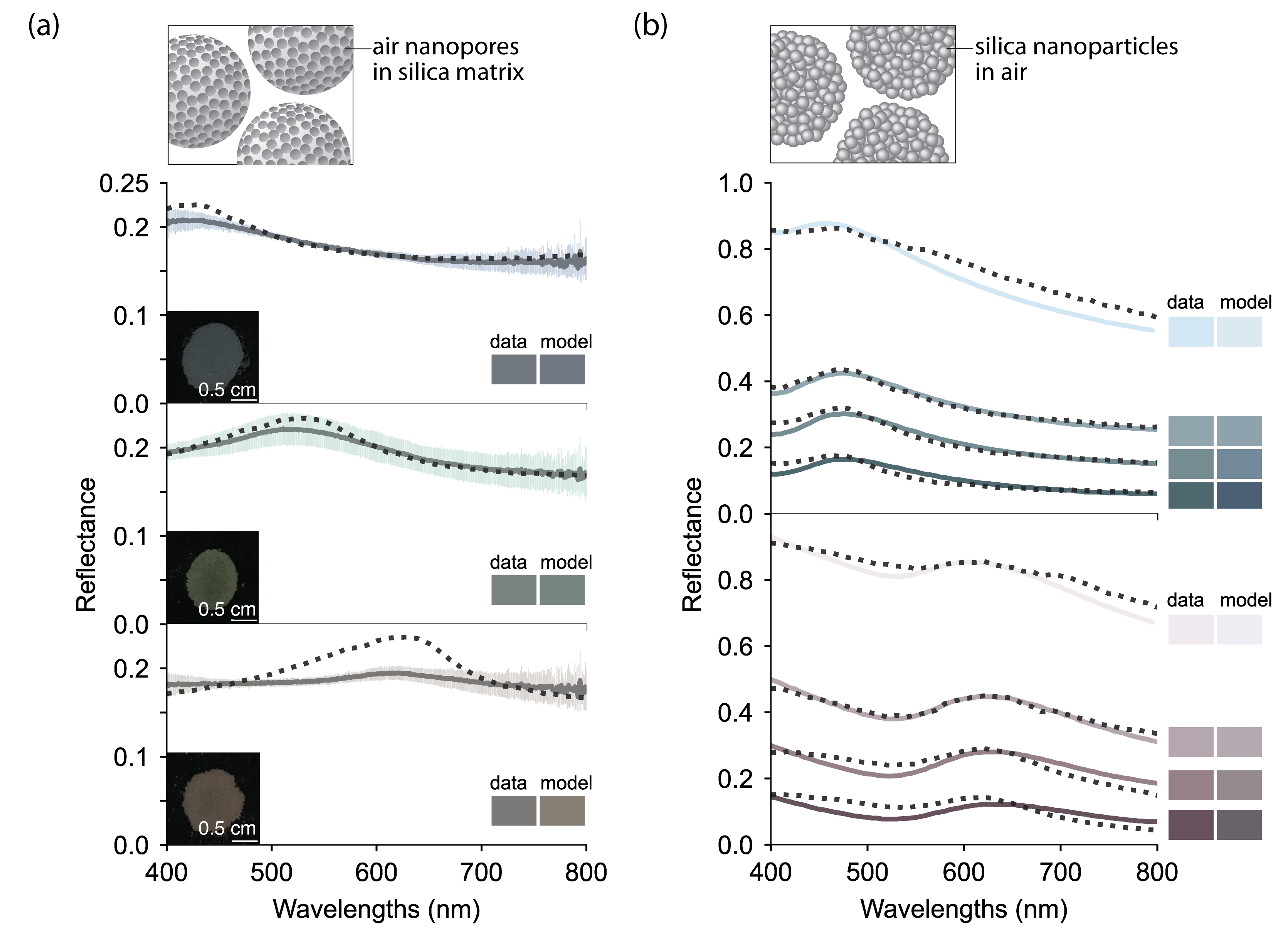}
  \caption{Measurements validate the multiscale Monte Carlo model for
    films of photonic balls. (a) Comparison of calculated (dotted lines) and measured
    (solid lines) reflectance spectra for films of inverse photonic balls consisting
    of nanopores in a sintered silica matrix. Model parameters for the
    blue (top), green (middle), red (bottom) samples, listed in order
    for each color: nanopore radius \SI{99}{\nm}, \SI{130}{\nm},
    \SI{163}{\nm}; nanopore volume fraction 0.5, 0.54, 0.6;
    polydispersity 0.0481, 0.0188, 0.0355; photonic-ball diameter
    \SI{18.5}{\um}, \SI{18.4}{\um}, \SI{19.6}{\um}; photonic-ball volume
    fraction 0.49, 0.5, 0.49; film thickness \SI{222}{\um},
    \SI{200}{\um}, \SI{200}{\um}; imaginary index of film matrix
    \SI{2.54e-3}{}, \SI{2.82e-3}{}, \SI{2.76e-3}{}. Model parameters
    used across all three samples: fine roughness 0.4, coarse roughness
    0.1, and the wavelength-dependent refractive index of the
    photonic-ball matrix~\cite{malitson_interspecimen_1965,
      kitamura_optical_2007}. (b) Comparison of calculated and measured
    reflectance spectra for films of direct photonic balls consisting of
    silica particles in an air matrix and containing varying
    concentrations of carbon black. The measured reflectance spectra are
    plotted from data in Ref.~\citenum{takeoka_production_2013}, which
    describes the fabrication and measurement procedures. For each color
    we use the following model parameters: thickness \SI{200}{\um},
    particle volume fraction 0.54, photonic-ball diameter \SI{7}{\um},
    photonic-ball volume fraction 0.55, polydispersity 0.05, coarse
    roughness 0.2. We use a fine roughness of 0.05 for each sample,
    except for the two that contain no carbon black, where we use a fine
    roughness of 0.01. For the green samples, we use a particle radius
    of \SI{109}{nm}. For the purple samples, we use a particle radius of
    \SI{144}{nm}. The imaginary refractive indices are varied for each
    sample in proportion to carbon-black concentration and are plotted
    in Fig.~S2.}
  \label{validation_pb_film}

\end{figure*}

The model produces good agreement for the magnitudes and peak positions
of the reflectance spectra for all three colors
(Fig.~\ref{validation_pb_film}a), and the predicted colors match the
measured colors well for the blue and green samples. We observe a
discrepancy between predicted and measured colors for the red sample.
This discrepancy might arise from differences between the structure that
is modeled and the actual structure of the sample. The observed
peak-to-background ratio of the red sample is lower than that of the
blue or green samples, suggesting weaker structural correlations and
differences in the nanopore structure of the red photonic balls compared
to the blue and green samples.

To further validate our model, we also compare our simulated spectra to
reflectance spectra of disordered photonic-ball films fabricated and
measured by Takeoka and colleagues~\cite{takeoka_production_2013}. This
comparison allows us to test the model on photonic-ball films that are
direct structures of nanoparticles, rather than inverse structures of
nanopores, and that are made using a different fabrication technique and
with a different placement of absorber. The photonic balls are composed
of silica nanoparticles in air and are packed into \SI{200}{\um}-thick
films. Using input parameters that are within the uncertainty of the
measured values (see Supporting Information), we find good agreement
between our multiscale model and the measured reflectance spectra, as
well as good agreement between the calculated and measured colors
(Fig.~\ref{validation_pb_film}b).

\begin{figure*} 
  \centering\includegraphics{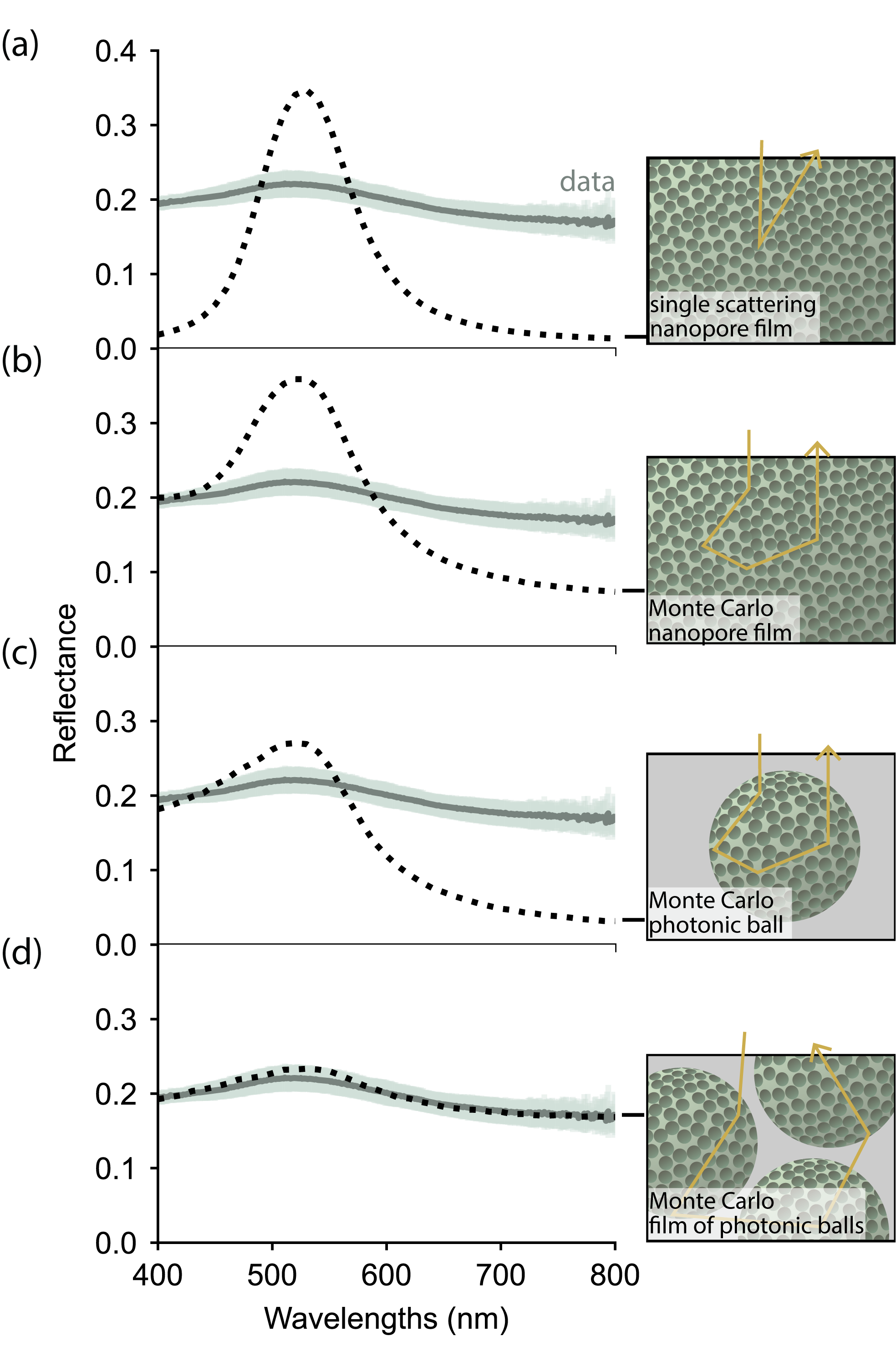}
  \caption{Comparison of measured (solid lines) and calculated (dotted lines) reflectance spectra
produced by different modeling approaches for a green film of photonic
balls. Measured reflectance data is for the green photonic-ball film shown in
Fig.~\ref{validation_pb_film}a. Calculated spectra are shown for (a)
single-scattering model for a nanostructured film, (b) Monte Carlo model
for a nanostructured film, (c) Monte Carlo model for an individual
photonic ball, and (d) multiscale Monte Carlo model for a film of
photonic balls.
Shared input parameters for each model:
    nanopore radius \SI{130}{\nm}, 
    nanopore volume fraction 0.54,
    nanopore polydispersity 0.0188.
For models with a photonic-ball film geometry, the film thickness is
\SI{220}{\um}. For the Monte Carlo model for a photonic-ball film, the
photonic-ball volume fraction is 0.5. For models with a nanopore film
geometry, the film thickness is multiplied by the volume fraction of
photonic balls so that the volume of scattering material is the same as
in the photonic-ball film models: thickness = \SI{200}{\um} $\times$ 0.5
= \SI{100}{\um}. For models with a photonic-ball geometry, the
photonic-ball diameter is \SI{18.4}{\um}. For the Monte Carlo models for
a nanopore film and photonic-ball films, the fine roughness is 0.4 and
coarse roughness is 0.1. For the single photonic-ball Monte Carlo model,
the fine roughness is 0.4 and coarse roughness is not included.
Imaginary refractive indices are chosen such that the total absorber
volume is constant across the nanopore film and photonic-ball film
geometries (see Supporting Information). The imaginary index of the
photonic-ball film matrix is
    \SI{2.82e-3}, and the imaginary index of the nanopore film matrix is
    \SI{3.06e-3}{}.
    \label{model_comparison}}
\end{figure*}

To understand whether the multiscale nature of our model is essential to
the agreement with experiment, we compare the measured spectra for a
green photonic-ball film to the spectra calculated from five different
models, including our multiscale model (Fig.~\ref{model_comparison}; see
Fig.~S1 for comparisons to other measurements). Each of the four other
models accounts for some, but not all, of the physical effects that the
multiscale model accounts for. The simplest model accounts for only
single scattering within a planar film
geometry~\cite{magkiriadou_absence_2014} using Mie theory, an
effective-medium approximation, and a structure factor. This
single-scattering film model reproduces the correct peak position near
\SI{550}{nm}, but the reflectance amplitudes are far from the
experimental spectrum (Fig.~\ref{model_comparison}a). A Monte Carlo
model that accounts for multiple scattering in a planar film produces
better agreement with the measured reflectance at low wavelengths, but
the peak reflectance is still much higher than the data, and the
long-wavelength reflectance much lower (Fig.~\ref{model_comparison}b).
An improved version of this model that accounts for the geometry of an
individual photonic ball causes the peak height to decrease, bringing
the reflectance closer to the data, but this model still produces
discrepancies at long wavelengths (Fig.~\ref{model_comparison}c). These
remaining discrepancies cannot be resolved by adjusting the roughness
parameters, because the peak shape and height are not significantly
affected by the roughness. The multiscale Monte Carlo model resolves the
discrepancies (Fig.~\ref{model_comparison}d). These results show that
modeling the multiple scattering due to the nanostructure, the
scattering from the boundary of the photonic balls, and the inter-ball
scattering are all necessary to capture the experimental reflectance
features.

% To investigate the
% effect of the inter-ball scattering, we also perform a Monte Carlo
% simulation for a film of photonic balls with a homogeneous refractive
% index, meaning their internal porous structure is not taken into
% account. The reflectance spectrum is very low across the visible
% wavelength range, indicating, unsurprisingly, that the scattering due to
% the internal structure of the balls is a significant contributor to
% reflectance.

%%%%%%
\subsection*{Developing design rules}
%%%%%%

\begin{figure*} [!htb]
\centering\includegraphics{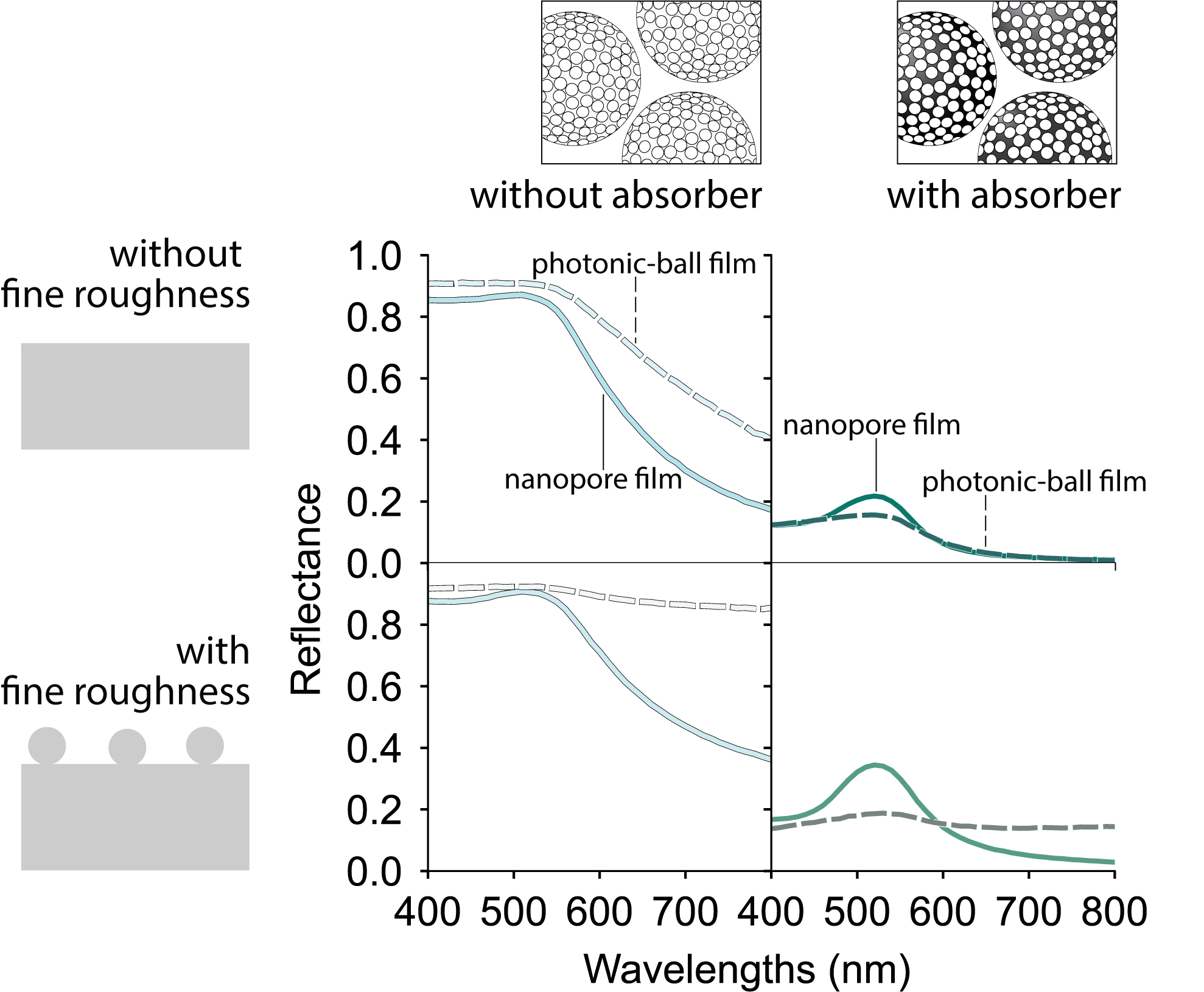}
\caption{Effects of fine roughness on the spectra of a photonic-ball
  film and nanopore film, as calculated by our Monte Carlo scheme.
  Without absorber and fine roughness, the reflectance of the nanopore
  film and photonic-ball film are similar. With fine roughness added,
  the reflectance of the photonic-ball film becomes flatter than that of
  the nanopore film, with more scattering at long wavelengths. Sample
  parameters are the same as those in Fig.~\ref{model_comparison} for
  the photonic-ball film and the nanopore film, except for panels
  showing no absorber and no fine roughness, where those values are set
  to zero. For the photonic-ball film, the absorber is added to the
  photonic-ball matrix. For the nanopore film, the absorber is added to
  the film matrix. In these simulations, the small amount 
  of measured absorption in silica is removed to allow for calculations
  with no absorption. The colors of the lines correspond to the color
  calculated from the spectrum using the CIELAB
  colorspace. \label{comparison_absorber_roughness}}
\end{figure*}

Having shown that the multiscale model can reproduce the spectral
features of photonic-ball films better than models that do not account
for inter- and intra-ball multiple scattering and the spherical
geometry, we now use the model to understand the physical mechanisms
behind spectral features. This understanding enables us to develop
design rules to make desired colors.

We first consider how and why the color of a photonic-ball film differs
from that of a film containing only nanopores with no higher-level
structure. The predicted reflectance spectrum of our green photonic-ball
film produces a much less saturated color than that of a comparable
nanopore film (Fig.~\ref{model_comparison}b), as evidenced by the high
off-peak scattering in the photonic-ball film spectrum. The reduced
saturation in the photonic-ball film could stem from the effects of fine
roughness or from where the broadband absorber is located. We use our
model to examine both cases.

In the absence of absorber and of fine roughness, the spectra of the
photonic-ball and nanopore films are similar, aside from the
photonic-ball film's slightly larger reflectance, which becomes more
pronounced at longer wavelengths (top left panel of
Fig.~\ref{comparison_absorber_roughness}). The larger reflectance likely
comes from Fresnel reflection and refraction at the boundaries of the
photonic balls. Light that reflects at the boundary can scatter more
times before it exits. The effect of the Fresnel reflections is more
prominent at longer wavelengths, where there is less scattering overall,
and therefore the relative increase in scattering is greater. Adding a
broadband absorber to both films while keeping the effective imaginary
indices equal reduces the overall reflectance in both (top right panel
of Fig.~\ref{comparison_absorber_roughness}). The reflectance peak of
the nanopore film becomes more pronounced, but the spectra are still
similar. These results show that the scattering in roughness-free films
is similar, whether they are composed of photonic balls or contain a
homogeneous arrangement of nanopores.

However, adding fine roughness leads to stark differences in the
reflectance spectra of the films. In the absence of absorber, adding
fine roughness raises the long-wavelength reflectance of the
photonic-ball film (bottom left panel of
Fig.~\ref{comparison_absorber_roughness}), leading to a much less
saturated color. When both fine roughness and absorption are included in
the model, the photonic-ball film shows a lower peak and nearly uniform,
broadband reflectance, while the nanopore film shows a more pronounced
peak with much lower long-wavelength reflectance
(bottom right panel of Fig.~\ref{comparison_absorber_roughness}).  Thus
the fine roughness is responsible for the difference in color between
photonic-ball films and homogeneous nanopore films.

The effects of fine roughness are magnified in a photonic-ball film
because of the higher surface area of the film compared to a homogeneous
nanopore film: each photonic ball has fine roughness, and therefore more
incoherent surface scattering occurs in the photonic-ball films. One way
to mitigate this effect---and achieve more saturated colors in
photonic-ball films---would be to reduce the fine roughness as much as
possible. However, this approach might require developing new
fabrication techniques. An alternative approach, which we consider next,
is to control the location of the absorber in the sample.

\begin{figure*}
\centering \includegraphics[width=\textwidth]{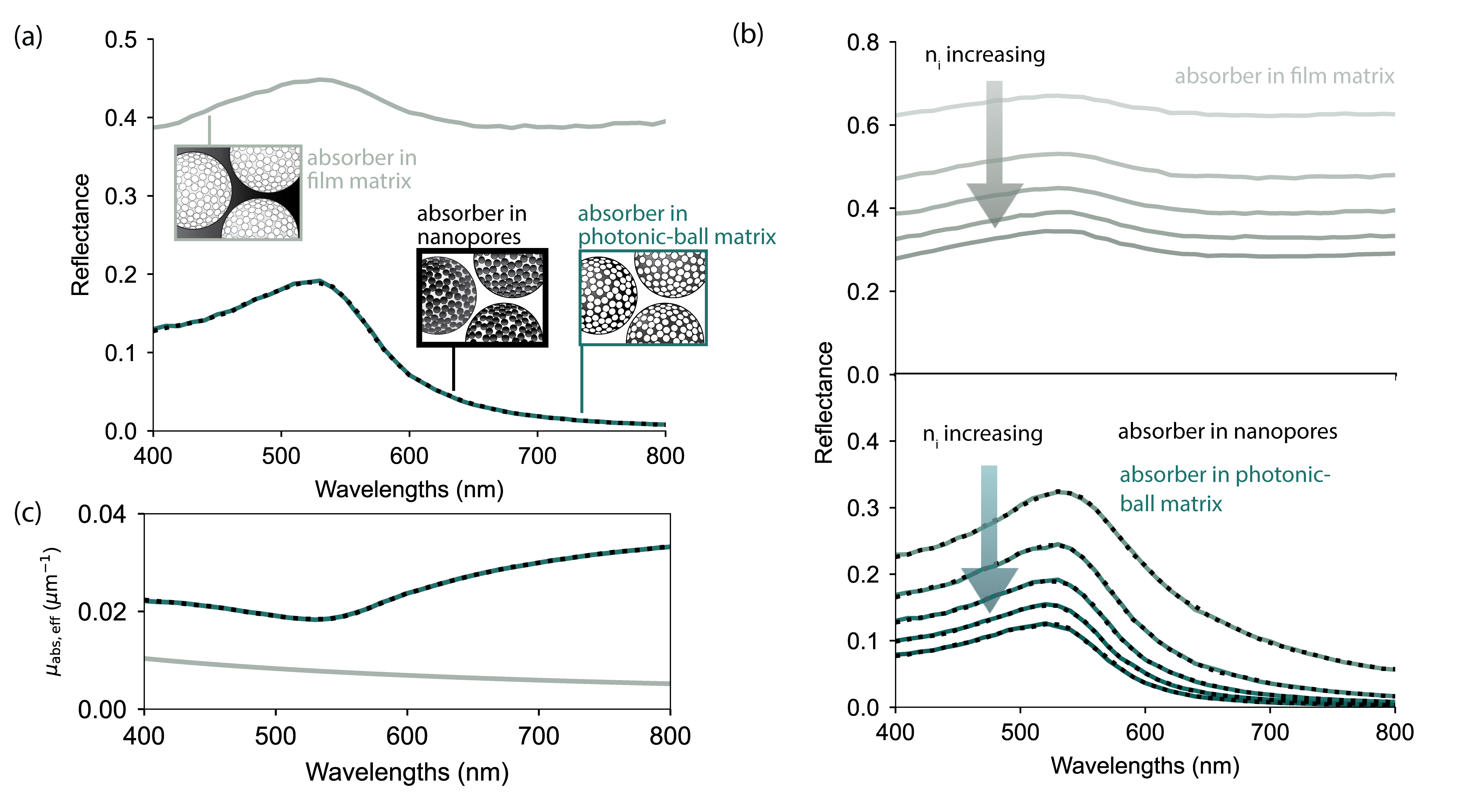}
\caption{Simulations show that placing absorber in the nanopores or
  photonic-ball matrix improves color saturation compared to placing
  absorber in the film matrix. Sample parameters are the same as those
  of the green photonic-ball film in Fig.~\ref{validation_pb_film}a, except
  for the imaginary refractive indices, which are listed below. (a)
  Reflectance spectra for three broadband absorber placements, each with
  equal effective imaginary indices for the sample. Insets show diagrams
  of the three absorber locations, with dark shaded regions indicating
  the absorber. The effective imaginary index for all three absorber
  placements is \SI{6.60e-4}{}. (b) Top: Reflectance spectra for a range
  of imaginary indices of the photonic-ball-film matrix. Bottom:
  Reflectance spectra for a range of imaginary indices of the
  photonic-ball matrix and nanopores. The effective imaginary indices
  used are \SI{8.25e-5}, \SI{2.06e-4}, \SI{3.30e-4}, \SI{4.53e-4}, and
  \SI{5.77e-4}{} (see SI for calculations of effective imaginary indices). 
  These values are the same for each of the three
  absorber locations to allow for comparison. (c) Effective absorption
  coefficient for the three different absorber locations as a function of
  wavelength. Effective imaginary index for each set of model parameters
  is \SI{3.30e-4}{}. \label{absorber_placement}}
\end{figure*}

Our multiscale simulations show that the color is much less saturated
when the absorber is placed in the film matrix compared to when it is in
the nanopores or in the photonic-ball matrix
(Fig.~\ref{absorber_placement}a). We also find that when the absorber
concentration is increased in the film matrix (top panel of
Fig.~\ref{absorber_placement}b), the entire reflectance spectrum shifts
downward. By contrast, when the absorber concentration is increased in
the photonic-ball matrix or nanopores, the long-wavelength, off-peak
reflectance decreases much more than the short-wavelength reflectance
does. The drop in long-wavelength reflectance leads to much more
saturated blue and green colors. Thus, placing the absorber in either
the nanopores or the photonic-ball matrix offers more saturated colors
and better control over the saturation. 

To understand why the off-peak reflectance decreases so much more than
the peak reflectance when the absorber is embedded inside the nanopores
or matrix of the photonic balls, we plot the effective absorption
coefficient for the photonic-ball film (Eq.~\eqref{abs_eff}) as a
function of wavelength (Fig.~\ref{absorber_placement}c). Although the
imaginary effective indices $n_i$ are equal for each film, the effective
absorption coefficient $\mu_{\textnormal{abs,eff}}$, which includes
absorption contributions from inside the photonic balls as well as from
the surrounding matrix, is greater when the absorber is in the nanopores
or matrix of the photonic balls. This increase in absorption occurs
because most of the scattering happens within the photonic balls, and so
placing the absorber inside the photonic balls reduces multiple
scattering to a greater extent than placing it outside the balls. The
multiple scattering is the principal contribution to the off-peak
reflectance. By contrast, there is less multiple scattering at the
reflectance peak, because much of the light is backscattered out of the
ball before it can scatter multiple times within the ball. As a result,
there is less absorption at the wavelength of the peak. This interplay
between scattering and absorption explains why placing the absorber
inside the photonic balls instead of the film matrix increases the
absorption preferentially at off-peak wavelengths, leading to more
saturated colors.

To further understand why there is little difference between placing the
absorber in the nanopores or in the photonic-ball matrix, we consider
how the effective imaginary index affects the reflectance. Because the
effective imaginary index is constrained to be equal for the two
placements (Fig.~\ref{absorber_placement}a), the absorption coefficient
$\mu_{\textnormal{abs}}$ calculated in the Monte Carlo simulation for
each individual photonic ball is also identical
(Fig.~\ref{absorber_placement}c). Therefore the intensity decreases with
distance in exactly the same way (set by Eq.~\eqref{beers_law_abs}) for
each ball. Although the imaginary index can in principle affect the form
factor $F$, calculated from the Mie solutions, the effects on $F$ are
negligible because the imaginary indices are small compared to the
difference between the real refractive indices. Thus the reflectance
values for absorber in the nanopores and in the photonic-ball matrix are
nearly identical. Note that the reflectances overlap only when the
imaginary indices are equal, even though the total volume of absorber
may be different across the samples. However, the reflectance trend of
increasing saturation remains the same whether absorber volume or
imaginary refractive index is increased.

\begin{figure*} 
\centering \includegraphics{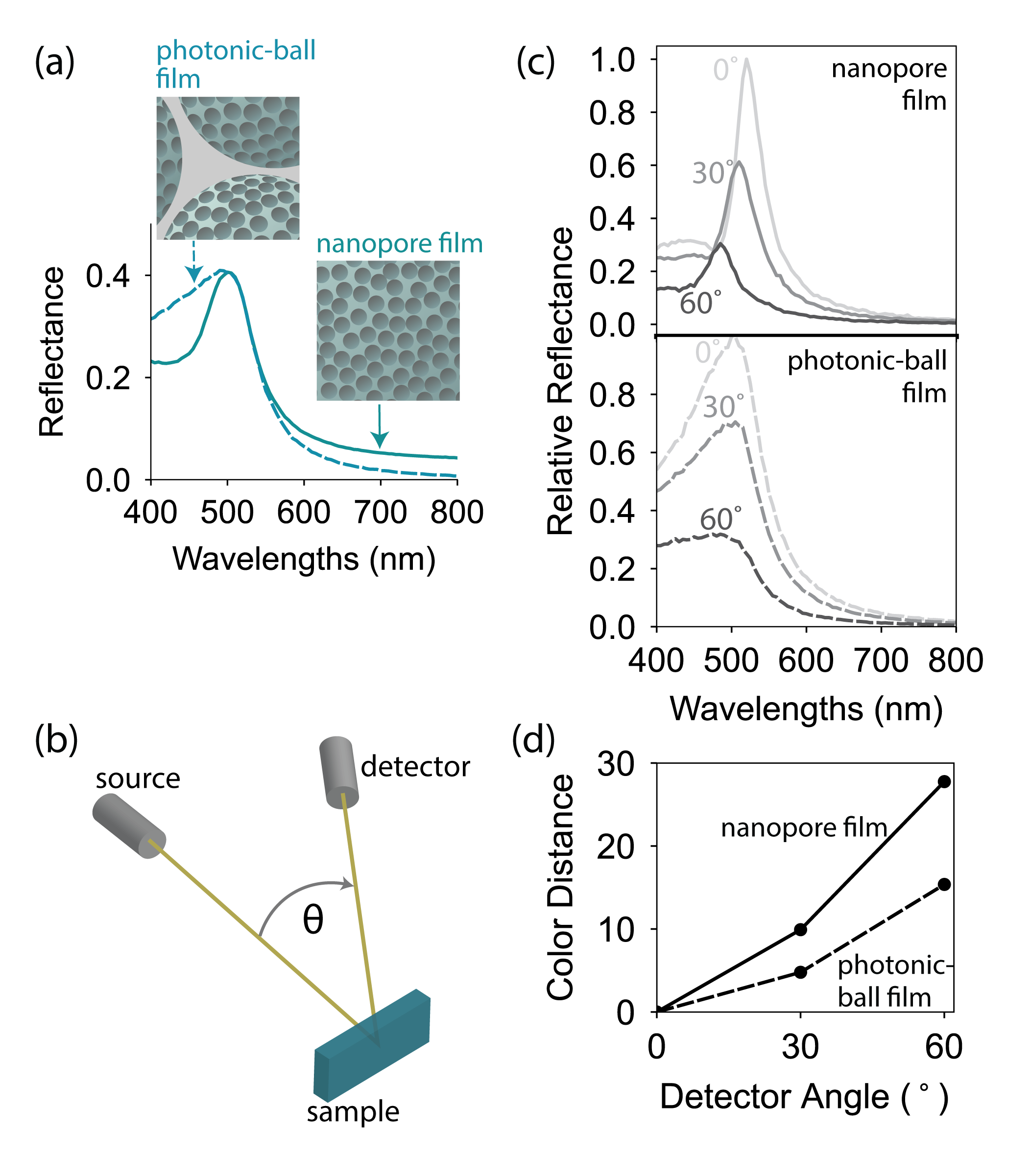}
\caption{Colors of photonic-ball films vary less with angle than do
  colors of homogeneous nanopore films. (a) Simulated reflectance
  spectra for a homogeneous nanopore film and a photonic-ball film, each
  with air pores and a silica matrix. Model parameters: nanopore radius
  \SI{130}{\nm}, pore volume fraction 0.6, fine roughness 0.15, coarse
  roughness 0.1. For the photonic-ball film, the volume fraction of
  photonic balls is 0.55, the photonic-ball diameter is \SI{10}{\um},
  and the thickness is \SI{200}{\um}. The nanopore-film thickness of
  \SI{110}{\um} is found by multiplying the photonic-ball film thickness
  by the photonic-ball volume fraction. The imaginary refractive indices
  are adjusted to make the heights of the reflectance peaks equal across
  the two simulations. For both the nanopore film and the photonic-ball
  film, the absorber is placed in the pores, with ${n_i}=\SI{6e-4}{}$
  for the nanopore film and ${n_i}=\SI{1.8e-4}{}$ for the photonic-ball
  film. (b) Diagram showing angle between source and detector. (c) Top:
  Simulated reflectance spectra for the nanopore film as a function of
  detection angle. Bottom: Simulated reflectance spectra for the
  photonic-ball film as a function of detection angle. We simulate a
  goniometer-style setup for a square detector \SI{10}{cm} from the
  sample, with a side length of \SI{5}{cm}, centered at the specified
  detection angle. Only photon packets whose paths intersect with the
  detector are counted. Reflectance is normalized to the highest
  reflectance value of the set of spectra to show a relative reflectance
  between 0 and 1. (d) CIE 1976 color difference comparison versus angle
  from the simulations shown in (b) and (c). \label{angle_dependence}}
\end{figure*}

Finally we examine the angle-dependence of the color, another important
property for applications. To characterize the angle-dependence of a
photonic-ball film, we simulate the reflectance for a range of detection
angles and compare the results to simulations for a nanopore film
(Fig.~\ref{angle_dependence}). To facilitate comparison, we simulate
spectra with similar colors (Fig.~\ref{angle_dependence}a) by choosing
the same base scattering parameters of material, nanopore size, volume
fraction, and roughness but varying the absorber concentration slightly
to match the colors. We find that both film geometries show the low
angle-dependence expected for disordered systems, but the photonic-ball
films show almost no shift in the peak of the reflectance with
wavelength (Fig.~\ref{angle_dependence}c). To quantify the color change
with angle, we calculate the CIE 1976 color difference
(Fig.~\ref{angle_dependence}d; see Supporting Information for
calculation details). The color of the photonic-ball film shifts much
less with angle than does the color of the nanopore film, and the small
color shift results from changes in the magnitude of the reflectance
rather than a change in the peak wavelength. This suppression of peak
shift suggests that photonic-ball films present a route to even more
angle-independent colors than can be achieved with nanostructured films,
making photonic-ball films promising for applications where a
homogeneous color is needed.

The increased angle-independence in a photonic-ball film can be
understood by considering the effect of the photonic-ball boundaries on
the incident light. When light is normally incident on the surface of a
photonic-ball film, it is refracted at the surface of each ball to a
different extent, depending on where it hits the ball surface (owing to
the curvature of the ball). Because of the variation in the direction of
light entering the ball, the reflectance at any given detection angle
will include light from a broad range of scattering angles. As shown in
a previous study\cite{xiao_investigating_2021}, a broader range of
scattering angles leads to less angle-dependence. The fundamental reason
is that the single-scattering interference that largely determines the
peak wavelength is a function of the scattering wavevector $|\textbf{q}|
= 4\pi \sin(\theta/2)/\lambda$, which couples the scattering angle
$\theta$ and wavelength $\lambda$.

%%%%%%%%%%%%%%
\section*{Conclusions}
%%%%%%%%%%%%%%

We have shown not only that a multiscale Monte Carlo model allows for
accurate predictions of the color of films of disordered photonic balls
with weak multiple scattering, but also that the model allows us to
understand how and why these colors differ from those of homogenous
nanostructured films. When we started our study, we hypothesized that
the two types of films would show similar reflectance spectra. Our
results show that this hypothesis is not correct for most photonic-ball
films that are made in practice. These films generally show reduced
color saturation compared to homogeneous nanostructured films. The
differences are related to exactly the effects that we account for in
our model: boundary effects at the photonic-ball surfaces and both
inter- and intra-ball multiple scattering. Specifically, we find that
refraction at the boundaries of the photonic balls leads to more
scattering at wavelengths longer than that of the reflectance peak.
Furthermore, because the surface area of a film of photonic balls is
much higher than that of a homogeneous nanostructured film, even small
amounts of surface roughness can markedly increase multiple scattering.
Both of these effects lead to reduced color saturation in photonic-balls
films. For those who want to make saturated structural colors from
photonic balls, these effects pose a challenge.

But we have also shown that it is possible to mitigate these effects by
controlling where absorber is placed in the film. In particular, placing
the absorber inside the nanopores or the matrix of the photonic balls
increases the saturation compared to placing it between the balls. This
increase in scattering arises because the film absorbs more multiply
scattered light when the absorber is placed inside the balls. Methods to
fabricate photonic balls could therefore aim to add the absorber
directly to the matrix precursor or to infiltrate the photonic balls
with absorber before the balls are packed together.

Although the color saturation of the resulting photonic-ball film may
still be lower than that of a homogenous nanostructured film (owing to
the effects of fine roughness), there is an advantage to using the
photonic-ball film: the angle-dependence should be weaker and the color
more uniform because of the refraction of light at the photonic-ball
surfaces. Though these disordered nanostructure assemblies offer more
angle-independence than their more ordered counterparts, the
photonic-ball film geometry further improves the angle-independence.
Thus, we argue that disordered photonic-ball films are indeed a
promising method to fabricate structural colors for applications, not
only because they can be readily added to coating and paint
formulations, but also because they have more angle-independent colors.

These design rules and our model open a path to engineering photonic
pigments for various applications. The model can be used as part of a
design process in which a target color is specified and an optimization
algorithm is used to determine the sample parameters (refractive
indices, particle size, volume fractions, and others) required to
achieve the best match to the target. To expand the palette of colors
that can be achieved, future work on fabrication methods might focus on
minimizing surface roughness, which would further increase saturation.
Future improvements to the model might include incorporating the exit
directions of totally internally reflected photon packets in the phase
function, calculating the scattering effects of absorbers such as carbon
black, and accounting for interference among different, multiply
scattered trajectories. Accounting for these physical effects may help
bring the model into even better agreement with measurements, thereby
enabling more accurate design of target colors. Furthermore, as shown in
previous work by Yazhgur and colleagues~\cite{yazhgur_scattering_2022},
there is a need to better understand the limits of index contrast where
effective-medium theories can be reliably used. Future work on this
subject could allow us to expand our model to study a wider range of
materials.

The model can readily be extended to account for scattering in other
hierarchical structures. Photonic balls containing nanoparticles or
nanopores in arrangements not described by the Percus-Yevick structure
factor---bicontinuous pore structures or packings of nanorods, for
example---can be modeled by first specifying the appropriate structure
factor, which can be derived or measured. Alternatively, if the angular
reflectance of the photonic ball can be measured or calculated, for
example through the finite-difference time-domain method, this data can
be used as input to the photonic-ball film model. These approaches would
allow modeling films of photonic balls containing a wide variety of
nanostructures, as long as these nanostructures are roughly isotropic.

\section*{Methods}
\label{sec:pb_methods}

\subsection*{Modeling light transport in a single photonic ball}

At the smallest scale of our model, we determine the single-scattering
properties from an assembly of nanoparticles or nanopores
(Fig.~\ref{model_overview}a). To calculate the scattering and absorption
from individual nanoparticles or nanopores, we treat the surrounding
nanoparticles or pores and the photonic-ball matrix material as a
homogeneous, effective medium. We use the Bruggeman
approximation~\cite{markel_introduction_2016} to calculate the effective
index of refraction from the volume fraction of nanoparticles, their
refractive index, and the refractive index of the photonic-ball matrix.
We then use Mie theory~\cite{bohren_absorption_2004} for a homogeneous
sphere embedded in the effective medium to calculate the form factor $F$
of the nanoparticle or nanopore, which is its differential scattering
cross-section as a function of angle. We use a structure factor $S$ to
take into account the constructive interference that results from the
short-range order in the packing of the particles. We assume the spheres
are packed as a hard-sphere liquid, such that $S$ can be described by
the Percus-Yevick approximation~\cite{ashcroft_structure_1966,
  xiao_investigating_2021}. The total scattering cross-section of the
sample $\sigma_{\textnormal{sca}}$ is
\begin{equation}
\sigma_{\textnormal{sca}} = \int \frac{d\sigma_{\textrm{sca}}}{d\Omega} d\Omega,
\end{equation}
where $d\sigma_{\textrm{sca}}/d\Omega$ is the differential scattering
cross-section for the nanostructure
packing~\cite{magkiriadou_absence_2014}: 
\begin{equation} \label{diff_cscat}
  \frac{d\sigma_{\textnormal{sca}}}{d\Omega} = FS.
\end{equation}
This combination of Mie theory, effective-medium theory, and structure
factor is the foundation of the single-scattering model developed by
Magkiriadou and colleagues to predict the reflectance from a film of
disordered colloids~\cite{magkiriadou_absence_2014} as well as the first
step of our multiscale model.

From these results we calculate two distributions. The first is the
distribution of distances (which we call steps) that light travels
before being scattered. This distribution is based on the Beer-Lambert law:
\begin{equation} \label{beers_law_scat}
  P(\textnormal{step}) =\frac{1}{l_{\textnormal{sca}}}\exp(-\text{step}/l_{\textnormal{sca}}).
\end{equation}
The mean of this distribution is the scattering length
$l_\textnormal{sca}=1/\rho\sigma_{\textnormal{sca}}$, where $\rho$
is the number density of nanoparticles. The second distribution is the
phase function, or the distribution of angles that the light scatters
into:
\begin{equation} \label{phase_fun} P(\theta) =
  \frac{d\sigma_{\textnormal{sca}}/d\Omega} {\sigma_{\textnormal{sca}}}.
\end{equation}

To simulate multiple scattering within a photonic ball, we use a Monte
Carlo scheme based on these two distributions
(Fig.~\ref{model_overview}b). We simulate trajectories of photon
``packets,'' which have a wavelength, initial direction, and initial
weight which can be reduced by absorption. The photon packets travel
linearly until they are scattered; we determine the distance they travel
by sampling from the step-size distribution, Eq.~\eqref{beers_law_scat}.
The scattering event changes the direction of the photons. We sample the
direction from the phase function, Eq.~\eqref{phase_fun}.

Because the materials used for the nanoparticles or matrix material may
have some absorption, and broadband absorbers such as carbon black are
often added to structurally colored materials, we must also account for
absorption in the model. When a component material has a complex
refractive index, we calculate the appropriate Mie solutions for
absorbing materials, as described by Hwang and
coworkers~\cite{hwang_designing_2021}. In addition, we calculate the
absorption length as
\begin{equation}
l_{\textnormal{abs}}=\frac{1}{\mu_{\textnormal{abs}}}=\frac{\lambda_0}{4\pi n_i},
\end{equation}
where $n_i$ is the imaginary component of the sample's effective
refractive index. By using this definition of absorption length, we are
assuming there are no strong resonances that would lead to higher
absorption than predicted by effective-medium theory. We then reduce the
weight of each photon packet according to the Beer-Lambert law:
\begin{equation} \label{beers_law_abs}
 W = W_i\exp(-\text{step}/l_{\textnormal{abs}}),
\end{equation}
where $W_i$ is the initial weight and $W$ the weight after the step. The
integrated reflectance is then calculated by summing the weights
that exit the sample in the reflection hemisphere.

As described thus far, this Monte Carlo scheme is the same as that of
Hwang and coworkers~\cite{hwang_designing_2021}, which yielded
reflectance spectra that were in quantitative agreement with
experimental measurements on disordered nanoparticle films. In applying
the scheme to a photonic ball, however, we must augment the model to
account for the boundary conditions of the ball. In particular, we
include refraction and Fresnel reflection at the boundary of the ball.
When a photon packet encounters the interface between the ball and its
surrounding medium, it splits into two packets, where one packet exits
the ball and the other is internally reflected back into the ball. The
weights of these two new packets are determined by the Fresnel
transmission and reflection coefficients, which in turn depend on the
effective index of the ball and the index of the matrix surrounding the
ball. The incident, refraction, and reflection angles are determined
with respect to the local normal vector on the photonic ball's surface.
We simulate the trajectory of the refracted packet as usual, then
simulate a new trajectory for the packet reflected back into the sphere.
When this packet eventually encounters the boundary, we again split it
and simulate a new trajectory for the reflected packet. This process is
repeated until the total weights of all packets left inside the ball is
less than \SI{1}{\percent} of the original packet weights. The details
of these calculations are discussed in the Supporting Information. 

\subsection*{Modeling light transport in a film of photonic balls}

Our scheme also differs from that of Hwang and
coworkers\cite{hwang_designing_2021} in that the aim is to extract a
phase function, total scattering cross-section, and total absorption
cross-section for the photonic ball, which are needed for modeling light
transport in a film of many photonic balls. We therefore simulate each
photon packet until it exits the ball or is almost completely absorbed.

To obtain the photonic ball phase function, we start by tracking where
each photon packet exits the ball's surface and its weight upon exiting.
We then use Gaussian kernel density estimation to construct the phase
function from these exit positions and weights (see Supporting
Information for details).

To calculate the step-size distribution and absorption length, we must
first calculate the total scattering and absorption cross-sections for a
single photonic ball. To calculate the scattering cross-section from the
scattered photon weights, we must introduce a scaling factor to assign
appropriate units and set a maximum value for the cross-section. The
total scattering cross-section for a single photonic ball is
\begin{equation}
  \sigma_{\textnormal{sca, ball}}= 2\pi r^2 \frac{\sum W_{\textnormal{sca}}}{\sum W_{\textnormal{inc}}},
\end{equation}
where $r$ is the radius of the photonic ball, $W_{\textnormal{sca}}$ is
the scattered weight of a photon packet, and $W_{\textnormal{inc}}$ is
the incident weight of a photon packet. The summations are over all
photon packets. The scaling factor of $2\pi r^2$ is the maximum
scattering cross-section of a sphere in the geometrical-optics
limit~\cite{berg_new_2011}. Setting the maximum scattering cross-section
to the geometrical-optics maximum is justified by the size of the
photonic balls, which are roughly 12--75 times the wavelength. We note
that this use of geometrical optics affects only the scaling of the
cross-section, and we do not use geometrical optics to determine the
photon-packet trajectories. The photonic ball absorption cross-section
is then
\begin{equation}
  \sigma_{\textnormal{abs, ball}}= \pi r^2\left(1-\frac{\sum W_{\textnormal{sca}}}{\sum W_{\textnormal{inc}}}\right),
\end{equation}
where the cross-section is scaled by the factor $\pi r^2$, the maximum
absorption cross-section in the geometrical optics
limit~\cite{berg_new_2011}.

Next, we simulate trajectories in the photonic-ball film by sampling
probability distributions for the step size and direction, using the
same method as in the first simulation but with different distributions
(Fig.~\ref{model_overview}b,c). The step size is sampled from the
distribution in Eq.~\eqref{beers_law_scat}, where $l_{\textnormal{sca}}$
is replaced with $l_{\textnormal{sca, ball}}$, the scattering length in
a photonic-ball film:
\begin{equation}
P(\textnormal{step}) = \frac{1}{l_{\textnormal{sca, ball}}}\exp(-\text{step}/l_{\textnormal{sca, ball}}),
\end{equation}
where
\begin{equation}
l_{\textnormal{sca, ball}} = \frac{1}{\rho_{\textnormal{ball}} \sigma_{\textnormal{sca, ball}}},
\end{equation}
and $\rho_{\textnormal{ball}}$ is the number density of photonic balls
inside the film. The direction is sampled from the photonic-ball phase
function, calculated from the exit positions and weights of the photon
packets as described above. To account for absorption, we decrease
the weights according to Eq.~\eqref{beers_law_abs}, replacing
$l_{\textnormal{abs}}$ with an effective absorption length for the
photonic-ball film:
\begin{equation} \label{abs_eff}
l_{\textnormal{abs, eff}} = \frac{1}{\mu_{\textnormal{abs, eff}}}, \\
\end{equation}
where $\mu_{\textnormal{abs, eff}}$ is the effective absorption
coefficient,
\begin{equation}
\mu_{\textnormal{abs, eff}} = \rho_{\textnormal{ball}} \sigma_{\textnormal{abs, ball}} + \frac{4 \pi n_i}{\lambda}(1-\phi_\textnormal{ball}),
\end{equation}
$\phi_\textnormal{ball}$ is the volume fraction of photonic balls in the
film, and $n_i$ is the imaginary refractive index of the matrix material
surrounding the photonic balls.

We calculate the integrated reflectance by adding the reflected weights
of trajectories and normalizing by the total incident weights. From the
reflectance, we can determine color coordinates for the perceived color
of the sample using the CIE color matching functions, as described by
Xiao and colleagues~\cite{xiao_investigating_2021}.

There are a few differences between the two Monte Carlo simulations due
to the different scales on which they operate. Because the photonic
balls are much larger than the wavelengths of visible light, we cannot
use effective-medium theory in the second level of our model. Instead we
consider the photonic balls to be embedded in the matrix material of the
film, and not in an effective medium. Also, because the photonic balls
are large compared to the wavelength, we neglect interference of light
scattered from different photonic balls. This assumption is equivalent
to choosing $S=1$ for the structure factor of the photonic-ball film.
Finally, we take into account surface roughness as described by Hwang
and coworkers~\cite{hwang_effects_2020} by considering two separate
scales of roughness: coarse and fine. The coarse roughness accounts for
sample roughness at a scale greater than the wavelength of light, which
changes the orientation of the sample surface at each point in space,
thereby modifying the refraction angles and Fresnel reflection
coefficients when photon packets hit the sample interface. We include in
our model a coarse roughness parameter, which is the root-mean-square
slope of the surface~\cite{van_ginneken_diffuse_1998}. The fine
roughness accounts for wavelength-scale roughness, produced by features
such as particles protruding from the sample surface. We model fine
roughness by excluding the structure-factor contribution from the
calculation of the step-size distribution for the first step of the
photon packets. The fine roughness parameter in our model is the
fraction of photon packets that encounter this roughness at the scale of
a single particle. We model coarse roughness only at the film surface
and fine roughness only at the photonic-ball surfaces, and for both
roughness types, we model the effect of roughness when light enters the
film or ball, but not when it exits.

\subsection*{Photonic ball fabrication}

To produce photonic balls of varying nanopore size, we first synthesize
sacrificial polymer templates. Three separate traditional emulsion
polymerizations were performed to produce poly(methyl meth\-ac\-ry\-late)
colloidal dispersions, each poly(methyl methacrylate) nanoparticle
having an average particle diameter of \SI{250}{\nm}, \SI{332}{\nm}, and
\SI{402}{\nm}.

The aqueous poly(methyl methacrylate) colloidal dispersion was diluted
to \SI{1}{\percent} w/w with deionized water containing \SI{1}{\percent}
w/w colloidal silica (Ludox SM colloidal silica, Sigma Aldrich). The
mixture was sonicated to prevent particle agglomeration. This aqueous
dispersion was injected by syringe pump into a microfluidic device
having a \SI{50}{\um} droplet junction. At the same time, a continuous
oil phase containing \SI{0.1}{\percent} w/w polyethylene
glycol/perfluoropolyether surfactant (FluoroSurfactant, RAN
Biotechnologies) in a fluorinated oil (Novec 7500, 3M) was injected into
the same device. When the device started to produce droplets, the
droplets were collected in a beaker containing fluorinated oil.

The collected droplets were then dried in an oven at \SI{45}{\celsius}
for \SI{4}{\hour} and then calcined by placing on a silicon wafer,
heating from room temperature to \SI{500}{\celsius} over \SI{4}{\hour},
holding at \SI{500}{\celsius} for \SI{2}{\hour}, and cooling back to
room temperature. This procedure results in photonic balls as a dried
powder. To produce photonic balls containing \SI{3}{\percent} w/w carbon
black, we mix \SI{10}{\milli\gram} of photonic balls with
\SI{150}{\milli\gram} of an aqueous dispersion of carbon black
(\SI{0.2}{\percent} w/w, Covarine Black WS 9199, Sensient Cosmetic
Technologies), and then dry at \SI{70}{\celsius} for \SI{1}{\hour}.

\subsection*{Photonic-ball film fabrication}

To make the structurally colored films shown in
Fig.~\ref{validation_pb_film}a, we make ordered photonic balls of three
different colors. These photonic balls are made using the procedure
described above and consist of a silica matrix with air nanopores. The
nanopore sizes are \SI{208\pm 10}{\nm} (blue), \SI{265 \pm 5}{\nm}
(green), \SI{338 \pm 12}{\nm} (red). We deposit photonic balls onto a
glass slide, then place a second glass slide on top of the deposited
photonic balls. We then compress the two glass slides on both sides
using binder clips and seal the edges of this sample chamber using 5-min
epoxy. After the epoxy is cured, we remove the binder clips.

\subsection*{Reflectance measurements}

We measure the reflectance spectra of films of photonic balls using an
Agilent Cary 7000 Universal Measurement Spectrophotometer with an
attached integrating sphere that collects the light scattered into the
reflection hemisphere. The sample is illuminated with light from a
double out-of-plane Littrow monochromator on a $\SI{1}{\milli\meter}
\times \SI{3}{\milli\meter}$ rectangular spot. The sample is placed
behind a circular port with a \SI{6}{\milli\meter} diameter. We
normalize the intensity measurement to a diffuse white reflectance
standard (Spectralon, Labsphere). We calculate error bars (shown in gray
in spectral data) as two standard deviations about the mean of
measurements taken at different locations on the samples. For the blue
sample, we measure 8 locations. For the green sample, we measure 11
locations. For the red sample, we measure 10 locations. We convert the
average reflectance spectra into a color swatch using the software
package ColorPy~\cite{kness_colorpy_nodate}.

\subsection*{Model}

The hierarchical Monte Carlo model was written in Python and is
available as an open-source package on
GitHub~\cite{magkiriadou_structural-color_nodate}.

\begin{suppinfo}

  Details of the model design and implementation, colorspace
  calculations, sample parameter measurement and estimation, absorber
  concentration calculations, additional results and validations, as
  well as a detailed discussion of the use of effective-medium theory in
  our model.
 
\end{suppinfo}

\begin{acknowledgement} 
  We thank Bernhard von Vacano, Rupert Konradi, Jennifer McGuire, and
  Audrey von Raesfeld for helpful discussions. We thank Martin Panchula
  for suggesting the multiscale modeling approach and Diane Tom for the
  SEM imaging. We thank Lukas Schertel and Geoffroy Aubry for providing
  their scattering strength data for comparison with our model. 
\end{acknowledgement}

\section*{Funding Sources}
Anna B. Stephenson acknowledges the support of the National Science Foundation
  (NSF) Graduate Research Fellowship Program under grant number
  DGE-1745303. This work was funded by BASF through the Northeast
  Research Alliance. This research was partially supported by NSF
  through the Harvard University Materials Research Science and
  Engineering Center under grant number DMR-2011754. In addition, this
  work was performed in part at the Center for Nanoscale Systems (CNS),
  a member of the National Nanotechnology Coordinated Infrastructure
  Network (NNCI), which is supported by the National Science Foundation
  under grant number EECS-1541959. CNS is part of Harvard University.

\bibliography{references}
\end{document}

Number of pages: 27

Number of figures: 5

Number of tables: 1

\pagebreak

%%%%%%%%%%%% 
\section{Model details} 
%%%%%%%%%%%% 

%%%%%%%%%%%%%%%%%%%%%%%%%%%%%%%%%% 
\subsection{Accounting for the spherical boundary of photonic balls}
%%%%%%%%%%%%%%%%%%%%%%%%%%%%%%%%%% 

In applying the modeling scheme from Hwang and
colleagues~\cite{hwang_designing_2021} to a photonic ball, we must
augment the model to account for the boundary conditions of the ball.

First, we must use a different algorithm to determine whether a photon
packet has left the photonic ball. Instead of checking their position in
$z$, which would tell us their depth in the film, we must check their
radial position from the center of the photonic ball. Photon packets
that have attempted to exit the photonic ball have trajectories that
satisfy the equation below at some time step:
\begin{equation} \label{potential_exits}
x^2 + y^2 + (z-\textnormal{radius})^2 > \textnormal{radius}^2,
\end{equation}
where $x$, $y$, and $z$ are the global positions of the photon packet,
and ``radius'' is the radius of the photonic ball. The subtraction of
the radius from $z$ is a result of the definition of our coordinate
system, where the center of the sphere is found at the point $x=0, y=0,
z=\textnormal{radius}$.

These photon packets are said to have attempted an exit because
satisfying this equation alone is not enough to determine that they have
exited. Owing to their angle of exit, they may have been totally
internally reflected back into the ball.

We check for this internal reflection by finding the angle between the
photon packet's trajectory and the vector normal to the ball surface and
applying Snell's law. To find the normal vector, we analytically solve
for the intersection between the line of the photon's trajectory and the
sphere surface at the photon packet exit. The solution is found by
solving a system of parametric equations. The equation for a
3-dimensional line in parametric form is given by:
\begin{equation}
\begin{aligned} \label{line}
 x &= x_0 + (x_1-x_0)t, \\
 y &= y_0 + (y_1-y_0)t, \\
 z &= z_0 + (z_1-z_0)t,
\end{aligned}
\end{equation}
where the line is parameterized by $t$, and $(x_0, y_0, z_0)$ is the
point on the photon packet's trajectory just before the attempted exit
takes place. The point $(x_1, y_1, z_1)$ is the photon packet's position
just after the attempted exit. Substituting this line equation into the
equation of a sphere gives the quadratic equation:
\begin{equation} 
dt^2 +ft +g =0,
\end{equation}
where
\begin{equation}
\begin{aligned} 
    d &= k_x^2 + k_y^2 + k_z^2, \\
    f &= 2(k_xx_0 + k_yy_0 + k_zz_0), \\
    g &= x_0^2 + y_0^2 + z_0^2-\textnormal{radius}^2, 
\end{aligned}
\end{equation}
where we define a vector $\textbf{k}$ for the trajectory's propagation
direction at the attempted exit,
\begin{equation} 
k = (x_1-x_0) + (y_1-y_0) + (z_1-z_0),
\end{equation}
and the solution for $t$ is
\begin{equation} 
t = (-f \pm \sqrt{f^2-4dg})/(2d). \\
\end{equation} 

We can express the trajectory-sphere intersection point as
\begin{equation}
\begin{aligned} 
x_{\textnormal{int}} &= x_0 + tk_x, \\
y_{\textnormal{int}} &= y_0 + tk_y, \\
z_{\textnormal{int}} &= z_0 + tk_z, 
\end{aligned}
\end{equation}
where $ (x_{\textnormal{int}}, y_{\textnormal{int}},
z_{\textnormal{int}} ) $ is the photon's exit point on the ball's
surface. Since $t$ has two solutions, leading to two intersection
points, we take only the point that is closest to the photon packet's
position after exit. We take the dot product between the photon packet's
exit vector and the normal vector to find the angle at which the photon
packet attempts to exit:
\begin{equation} 
n_1\sin \theta_1=n_2\sin \theta_2,
\end{equation} 
where $n_1$ is the effective refractive index of the photonic ball, and
$n_2$ is the effective refractive index of the surrounding matrix, and
$\theta_1$ and $\theta_2$ are the angles from the normal before and
after refraction.

Many photon packets, however, are not totally internally reflected, and
at least some portion of their weight exits the photonic ball. When
calculating the reflectance of an individual photonic ball, we must
determine which photon packets exit into the reflection hemisphere. This
requires that we accurately calculate the photon packet direction after
exit, so we must include the direction change due to refraction by
rotating $\textbf{k}$, the vector that describes the propagation
direction. We perform this rotation using a matrix that rotates a vector
about an arbitrary axis~\cite{kovacs_rotation_2012}. The axis about
which we rotate $\textbf{k}$ is the cross product of $\textbf{k}$ and
the normal vector of the sphere surface at the photon packet's exit
position. We refer to $\textbf{k}$ before rotation as $\left<k_{x,1},
  k_{y,1}, k_{z,1}\right>$ and the exit point on the ball's surface as
$(a,b,c)$. The angle about which we are rotating is $\alpha = -
(\theta_2 - \theta_1)$.

To perform this rotation, we must first express a point on the line in
the direction of the initial $\textbf{k}$:
\begin{equation}
\begin{aligned} 
x &= a + k_{x,1}, \\
y &= b + k_{y,1}, \\
z &= c + k_{z,1}. 
\end{aligned} 
\end{equation}
Then, we can perform the rotation, multiplying the vector $\left<x, y,
  z\right>$ by the rotation matrix:
\begin{equation}
\begin{aligned} 
x_\textnormal{rot} &= (a(v^2 + w^2)-u(bv+cw-ux-vy-wz))(1-\cos(\alpha)) \\&\qquad + x\cos(\alpha) + (-cv + bw - wy + vz)\sin(\alpha), \\
y_\textnormal{rot} &= (b(u^2 + w^2)-v(au+cw-ux-vy-wz))(1-\cos(\alpha)) \\&\qquad + y\cos(\alpha) + (cu - aw + wx - uz)\sin(\alpha), \\
  z_\textnormal{rot} &= (c(u^2 + v^2)-w(au+bv-ux-vy-wz))(1-\cos(\alpha)) \\&\qquad + z\cos(\alpha) + (-bu + av - vx + uy)\sin(\alpha), 
\end{aligned}
\end{equation}
where $\left<u,v,w\right>$ is the unit direction vector of the normal at
the exit point. Then we convert back to a vector:
\begin{equation}
\begin{aligned} 
k_{x,2} &= x_\textnormal{rot} - a, \\
k_{y,2} &= y_\textnormal{rot} - b, \\
k_{z,2} &= z_\textnormal{rot} - c, \\
\end{aligned}
\end{equation}
where $\left<k_{x,2}, k_{y,2}, k_{z,2}\right>$ is the rotated $\textbf{k}$.

We also calculate the Fresnel reflection at the entrance and exit. For
both the entrance and exit, we use the method described above to
calculate the angle between a photon packet's trajectory and the ball's
normal vector at the position of entrance into the ball or exit from the
ball. Using the angle of incidence, we then calculate the Fresnel
coefficients to determine the fraction of photon-packet weight that
successfully enters the photonic ball in the first time step, as well as
the fraction of photon packet weight that is internally reflected back
into the photonic ball upon an attempted exit. Some of these photons are
totally internally reflected, meaning that their full weight is
reflected back into the photonic ball.

%%%%%%%%%%%%%%%%%%%%%%%%%%%%%%%%%%%%%%%%%%
 \subsection{Splitting photon packets at the ball boundary}
 %%%%%%%%%%%%%%%%%%%%%%%%%%%%%%%%%%%%%%%%%%

 Finally, we have to handle the photon packets that are partially or
 totally internally reflected upon an attempted exit, as well as any
 photon packets that are still scattering inside the ball. We split
 packets into two components when they exit, where the weights of these
 packets are assigned according to their corresponding Fresnel
 coefficient. We then use a recursive process to simulate the scattering
 of these new photon packets. The reflected photon packet directions are
 calculated by performing a reflection of $\textbf{k}$ off of the plane
 tangent to the sphere surface at the position of the photon packet
 exit:
\begin{equation}
\textbf{k}_r = \textbf{k}_1 - 2(\textbf{k}_1 \cdot \hat{\textbf{n}})\hat{\textbf{n}}.
\end{equation}

The Monte Carlo simulation steps described in the text are repeated, and
we recursively simulate new trajectories within each simulation run
until the total photon weights inside the sample is no larger than 1\%
of the original weights. The remaining weights are distributed equally
between reflection and transmission:
\begin{equation}
\begin{aligned} 
R_{\textnormal{extra}} = 0.5W_{\textnormal{stuck}}, \\
T_{\textnormal{extra}} = 0.5W_{\textnormal{stuck}}, 
\end{aligned}
\end{equation} 
where $R_{\textnormal{extra}}$ is the extra reflectance,
$T_{\textnormal{extra}} $ is the extra transmittance, and
$W_{\textnormal{stuck}}$ is the normalized sum of the trajectory
weights stuck inside the photonic ball.

Using this method, we calculate the total reflectance and transmittance
for a single photonic ball, as well as the reflectance per original
photon packet and the transmittance per original photon packet.

%%%%%%%%%%%%%%%%%%%%%%%%%%%%%%%%%%%
\subsection{Photonic-ball phase function calculation}
%%%%%%%%%%%%%%%%%%%%%%%%%%%%%%%%%%%

To calculate the distribution of directions, or \textit{phase function},
for the photonic ball, we first assign an angle to each photon packet by
finding the position on the ball's surface where each photon packet
exits. In our recursive method of calculating the reflectance and
transmittance per photon packet, we don't track the exit positions of
all of the additional photon packets that are split from the initial
photon packet. Instead, we add up the exit weights of any photon packets
that derived from one original packet and associate that full weight
with the exit position of the originating photon. The assumption behind
this procedure is that the distribution of initial exit positions is
representative of the distributions of exits of the photon packets
produced from the splitting at the boundary.

In assigning an angle to each photon packet, we use the angle of the
exit position on the surface rather than the exit direction, which
enforces an assumption that the photon packets exit normal to the sphere
surface. Though the simulated photon packets do not necessarily exit the
ball normal to the surface, the alternative of using the exit directions
would neglect how exit positions can geometrically restrict the
direction of the next scattering event. Because unpolarized scattering
depends only on the scattering angle $\theta$ and not on the azimuthal
angle $\phi$, we assume that the distribution of azimuthal angles is
uniform, and we restrict our calculation of the photonic-ball phase
function to a distribution of $\theta$ only.

We use kernel density estimation to produce a probability distribution
of photon-packet directions. This distribution is the photonic-ball
phase function. We use Silverman's rule for unequally weighted data to
select the kernel bandwidth~\cite{silverman_density_1986}:
\begin{equation}
  \left( \frac{3m_{\textnormal{eff}}}{4} \right)^{-1/5},
\end{equation}
where $m_{\textnormal{eff}}$ is the effective number of data points, defined as
\begin{equation}
  m_{\textnormal{eff}} = \frac{(\sum w_{\textnormal{sca}})^2}{\sum w_{\textnormal{sca}}^2}, 
 \end{equation}
 where $w_{\textnormal{sca}}$ is the weight of a scattered trajectory
 at exit. 

%%%%%%%%%%%%%%%%%%%%%%%%%%%%%%%%%%%
\subsection{Computational considerations for the Monte Carlo simulations} 
%%%%%%%%%%%%%%%%%%%%%%%%%%%%%%%%%%%

We run each integrated reflectance Monte Carlo simulation for 80,000
trajectories at each wavelength. For film geometries, we run the
simulations for 800 events. For photonic-ball geometries, we run the
simulations for an initial 300 events. We report the initial event
number, since the simulation splits an exiting photon packet into two
and continues to simulate the scattering trajectory of the internally
reflected photon packet for another 300 events. These nested simulations
run until the total weight of trajectories left inside the ball is less
than 1\% of the total initial weights. For the reflectance simulations
in which we detect scattered light over a small range of angles, we
increase the trajectory number to 300,000 since only a small fraction of
these trajectories encounter the detector.

We characterize the uncertainty by analyzing the results of 5 runs using
the parameters of our green photonic-ball-film sample. We calculate the
total weight percent of trajectories left inside the photonic-ball films
after 800 events, finding a mean of \SI{9.78e-6}{\percent} and standard
deviation \SI{2.06e-5}{\percent} across the 41 wavelengths and 5 runs.
To calculate an uncertainty for the reflectance values, we calculate the
standard deviation in reflectance across the 5 runs at each wavelength,
and then take the mean across the spectrum. This spectrum-averaged
standard deviation for the green photonic-ball film is
\SI{0.0775}{\percent}, well below the measurement uncertainty in our
samples. Performing the same calculation for the photonic balls that
compose the film yields a spectrum-averaged standard deviation of
\SI{0.127}{\percent}.

These simulations are run on Harvard's Cannon high-performance computing
cluster, on a single core using a maximum memory of \SI{119.4}{GB} with
an average clock speed of \SI{1.1}{GHz}. A typical simulation with the
above simulation parameters and these computational parameters takes
about \SI{12}{\minute} per wavelength. We choose the trajectory number
of 80,000 and the large limit of requested CPU memory to lower
uncertainty values while maximizing simulation speed. However, using a
high-performance computing cluster and such high trajectory-event
numbers is not necessary to obtain spectra with uncertainty less than
measurement uncertainty. To demonstrate this point, we also run quicker
simulations with the parameters of our green photonic-ball-film sample.
We run each simulation for 1,000 trajectories and 100 events for
photonic-ball and film geometries. For the individual photonic-ball
simulation, the total weights of stuck photon packets is still less than
1\% of the incident photon packet weights, because this number is a set
threshold. The mean weight percentages across 41 wavelengths and 5 runs
of photon packets left inside the photonic-ball films after 100 events
is \SI{5.75e-6}{} $ \pm $ \SI{ 2.9e-5}{\percent}. These percentages are
within the same order of magnitude as for the case of 800 events,
suggesting that for these sample parameters, 100 events is sufficient,
and increasing the event number to 800 offers no clear improvement. The
spectrum-averaged standard deviation in photonic-ball-film reflectance
across 5 runs at each wavelength is \SI{.705}{\percent}, and the
spectrum-averaged standard deviation in photonic-ball reflectance is
\SI{1.01}{\percent}. Running on a \SI{2.4}{\giga\hertz} core on a
machine with \SI{8}{\gibi\byte} of RAM, the simulations took a mean of
\SI{2.24}{\second} per wavelength.

%%%%%%%%%%%%%%%%%%%%%
\section{CIELAB colorspace calculations}
%%%%%%%%%%%%%%%%%%%%%
 
From our spectra, we calculate color coordinates that can then be used
to display colors on a screen or calculate the CIE 1976 color
difference, which characterizes the perceived differences between
colors. We calculate the CIELAB color coordinates $L^*$, $a^*$, and
$b^*$ and the RGB color coordinates using the method described by Xiao
and colleagues~\cite{xiao_investigating_2021}, where the calculations
are performed using the ColorPy Python
package~\cite{kness_colorpy_nodate}. The CIE 1976 color difference is
calculated as~\cite{klein_industrial_2010}
\begin{equation}
\Delta = \sqrt{(L^*_1 - L^*_0)^2 + (a^*_1-a^*_0)^2 + (b^*_1-b^*_0)^2},
\end{equation} 
where $(L^*_0, a^*_0, b^*_0)$ and $(L^*_1, a^*_1, b^*_1)$ are
coordinates for two colors.

%%%%%%%%%%%%%%%%%%%%%%%%%%%%% 
\section{Sample parameter measurement and estimation}
%%%%%%%%%%%%%%%%%%%%%%%%%%%%%

\subsection{Photonic ball parameter measurement and estimation}
\label{ball-parameter-estimation}

We perform SEM imaging to characterize the photonic balls and to
determine the average photonic-ball diameter and nanopore diameter.
Samples of each photonic ball size are loaded onto a conductive carbon
tape and sputtered with a platinum layer of \SI{1}{\nano\meter} in
thickness prior to imaging. The images are obtained on a JEOL cold
cathode Field Emission Scanning Electron Microscope in low detector LEI
mode. We measure the mean photonic-ball diameter and polydispersity
using ImageJ image analysis software, where 50--70 photonic balls are
measured. The mean nanopore diameter and standard deviation are measured
similarly by analyzing 50--70 nanopores across several photonic balls.
For the blue photonic balls, made with a \SI{250}{\nano\meter}
poly(methyl methacrylate) template, we measure a mean nanopore diameter
of \SI{208(10)}{\nano\meter} and a mean photonic ball diameter of
\SI{16.1(24)}{\micro\meter}. For the green photonic balls, made with a
\SI{332}{\nano\meter} poly(methyl methacrylate) template, we measure a
mean nanopore diameter of \SI{265(5)}{\nano\meter} and a mean photonic
ball diameter of \SI{17.8(27)}{\micro\meter}. For the red photonic
balls, made with a \SI{402}{\nano\meter} poly(methyl methacrylate)
template, we measure a mean nanopore diameter of
\SI{338(12)}{\nano\meter} and a mean photonic ball diameter of
\SI{16.3(33)}{\micro\meter}. We adjust the nanopore and photonic-ball
diameters that we input into the model within a standard deviation of
the mean to fit the model to the data.

Because nanopore volume fraction is difficult to measure using SEM, we
estimate the volume fraction within a range of reasonable values.
Collective jamming of hard spheres is thought to occur at volume
fractions as low as around 0.49, and the maximum randomly jammed state
occurs at a volume fraction around 0.64~\cite{torquato_jammed_2010}. We
therefore restrict the nanopore volume fraction estimates to this range.
The nanopore volume fraction primarily affects the reflectance peak
width and, to a certain extent, the peak position. We therefore
adjust the volume fraction within the range to improve agreement with
data in the reflectance peak width and position.

The photonic balls in our samples have a matrix of silica. We estimate
the real part of the refractive index using the Sellmeier dispersion
formula with parameters that were fit to experimental data for fused
silica~\cite{malitson_interspecimen_1965}. Measurements of the imaginary
refractive index of fused silica in the visible wavelength range yield
values ranging from roughly
\SIrange{7e-8}{1e-7}{}~\cite{kitamura_optical_2007}. For simplicity, we
use the midpoint between these estimated limits, \SI{8.5e-8}{}. Using
such an estimate is reasonable since these values are several orders of
magnitude lower than the imaginary index contribution from the carbon
black in our samples. Because the nanopores are assumed to contain only
air, we use a real refractive index of 1.

The fine roughness in our model accounts for wavelength-scale roughness
on the sample surface which leads to a breakdown in effective-medium
theory at the sample interface. Because of its small scale, the fine
roughness is difficult to measure, and we therefore adjust the fine
roughness parameter between 0 and 1 to best fit the data. However, we do
not fit the values for each individual sample. Instead, we fit the
values to all samples fabricated with the same method under the same
conditions, since we expect the fine roughness values to depend largely
on the drying and packing processes that occur during sample assembly.
Because the fine roughness primarily affects the magnitude of the
off-peak reflectance, we adjust the value to produce good agreement
between the predicted and measured off-peak reflectance magnitudes. This
method of empirically determining the roughness parameter for a given
sample fabrication protocol can then be used to design specific colors
made with that protocol.

When the sample preparation protocol is not known, we have no means to
determine the roughness parameters. In future work, we could address
this problem by using a more comprehensive model for the transition of
refractive index from the medium to the sample, such as the one
developed by Han and coworkers~\cite{han_determination_2019}.

\subsection{Photonic-ball film parameter measurement and estimation}
\label{ball-film-parameter-estimation}

We also measure or estimate the photonic-ball film properties that are
input into our model. To measure the thickness of the photonic-ball
film, we first use a micrometer to measure the thickness of the two
glass slides used to make the sample chamber. After preparing the film
sample, we use a micrometer to measure the total thickness of the sample
including the glass slides. We perform each thickness measurement 3--5
times across the sample surface. To find the thickness of the
photonic-ball film, we subtract the thickness of the two glass slides
from the total sample thickness. We find thicknesses of
\SI{227(5)}{\micro\meter} for the blue film, \SI{200}{\micro\meter} for
the green film, and \SI{195(5)}{\micro\meter} for the red film. We
adjust the parameter values used in the simulations within a standard
deviation about the mean.

Since the photonic-ball film is held in place by compression between two
glass slides, we assume the photonic balls are in a jammed packing in
the photonic-ball film. To find the photonic-ball volume fraction, we
therefore restrict our inputs to a range of reasonable values as
described in Section~\ref{ball-parameter-estimation}. Since the
photonic-ball volume fraction primarily affects the predicted
reflectance magnitudes across the spectrum, we use a value that results
in a broadband reflectance magnitude close to that of the data.

To obtain the refractive index of the matrix surrounding the photonic
balls, we perform a calculation based on the carbon black and silica
parameters. Owing to the small weight of carbon black added
(\SI{3}{\percent} w/w), we ignore the contribution of the real
refractive index of carbon black. We assume that the real component of
the matrix refractive index is 1, since the photonic balls are
surrounded by only air and carbon black. The carbon black used is a
suspension of carbon black nanoparticles in a water and glycerin mixture
(Covarine Black WS 9199, Sensient Cosmetic Technologies). The reported
density of the carbon black in the suspension is \SI{1.7}{g/mL}, the
reported size is \SIrange{1}{100}{\nm}, and the reported concentration
is \SI{25}{\percent} w/w. We use an imaginary index of 0.44 for carbon
black, which is commonly used in the
literature~\cite{dalmeida_atmospheric_1991, chylek_effect_1995}. We use
these values, combined with the volume fraction of the film matrix, the
total film weight, and total film volume to calculate the volume
fraction of carbon black in the film matrix. We then multiply this
carbon black volume fraction by the imaginary index of carbon black
(0.44) to obtain the imaginary index of the photonic-ball-film matrix.

We also specify the incident illumination and detection angles in our
model. The incident illumination angle is the angle between the light
source and the vector normal to the film plane. We use a value of
\ang{8}, which is the angle of the sample port of the integrating
sphere. We assume that the entire reflection hemisphere is captured by
the detector, as is expected for integrating sphere measurements. We
therefore set the detection angle range to \ang{90}--\ang{180}. Because
our films are between two glass slides, we also account for the Fresnel
reflections due to air-glass and sample-glass interfaces, using a
measured wavelength-dependent refractive index for soda-lime
glass~\cite{rubin_optical_1985}.

Because the coarse roughness parameter slightly affects the magnitude of
the reflectance across the spectrum, we set the value to produce good
agreement between the predicted and measured reflectance magnitudes. We
use the same coarse roughness parameter for our three samples, since
each film is fabricated using the same technique, and we expect the
sample interface shape to depend largely on the pressure and shape of
the interface enforced by the glass slides.

\subsection{Estimation of parameters of samples from literature}

We generate predicted reflectance spectra for the photonic balls from
Ref.~\citenum{zhao_angular-independent_2020} (Fig.~3 of the main text), 
starting with input parameters equal to the reported measured values for
nanopore size, photonic-ball size, and matrix refractive index. We
adjust the nanopore and photonic-ball diameters to achieve agreement
with reflectance peak positions in the data, and our parameters fall
within two standard deviations of the measured values. The photonic ball
consists of air nanopores embedded in a matrix made from an amphiphilic
bottlebrush block copolymer, P(PS-NB)-\textit{b}-P(PEO-NB). We use the
estimated refractive index of 1.52 reported by Zhao and
colleagues~\cite{zhao_angular-independent_2020}, which does not include
dispersion. We adjust the volume fraction using the same method
described in Sections~\ref{ball-parameter-estimation}
and~\ref{ball-film-parameter-estimation}. We adjust the fine roughness
value, which primarily affects the off-peak reflectance, to reproduce
the measured off-peak reflectance values, using the same fine roughness
value (0.01) for the three samples. Because these measurements are taken
through a microscope, the reflectance measured does not capture the
entire reflection hemisphere. An angular detection range of \ang{120}
results in good agreement with the data. The illumination angle is set
according to the experimental details of the setup, where the
illumination angle is limited by the numerical aperture of the objective
used (Zeiss, W N-Achroplan, 63$\times$, NA 0.9). Since the objective is
immersed in water, this NA gives a maximum illumination angle of
\ang{43}.

For the photonic-ball films from Ref.~\citenum{takeoka_production_2013}
(Fig.~4b of the main text), we adjust input parameters within reasonable 
uncertainty estimates of measured values. The reported hydrodynamic
diameters from dynamic light scattering are \SI{280}{\nano\meter}
(green) and \SI{360}{\nano\meter} (red). We use \SI{218}{\nano\meter}
and \SI{288}{\nano\meter} in our model, which are within 25\% of the
mean hydrodynamic diameters. Smaller values for optical diameters are
expected, because hydrodynamic diameters are often larger than optical
diameters. In these samples, carbon black is added during the photonic
ball assembly. Therefore, some of the carbon black may be located inside
the photonic balls, and some in the film matrix, outside the photonic
balls. We treat the sample imaginary index and the relative amounts of
carbon black inside and outside the photonic ball as fitting parameters.
The best-fit carbon black concentrations are proportional to carbon
black concentration, which indicates they are physically reasonable
(Fig.~\ref{S2}). We use an illumination angle of \ang{8}, as specified
by the authors. Since these measurements are taken using an integrating
sphere, we include reflectance contributions from the entire reflection
hemisphere. We also simulate the Fresnel reflections at the interface of
the substrate, a glass slide, using a measured wavelength-dependent
refractive index for soda-lime glass~\cite{rubin_optical_1985}. Because
the same assembly techniques are used across the eight samples shown
(Fig.~4b of the main text), we use the same coarse roughness parameter 
for each sample. We use the same fine roughness parameters for all of
the samples, except for the two samples that do not contain carbon
black. Because the presence of the carbon black nanoparticles could
influence the packing behavior of the silica nanoparticles, a different
fine roughness is expected for these samples.

%%%%%%%%%%%%%%%%%%%%%%%%%%%%% 
\section{Calculations involving absorbers} 
%%%%%%%%%%%%%%%%%%%%%%%%%%%%%

\subsection{Absorber volume in samples}

For the simulations shown in Fig.~5 of the main text, we keep the total
volume of carbon black constant across the simulations for the nanopore
film and the photonic-ball film. We choose to keep the carbon black
volume constant, rather than the effective imaginary refractive index,
because the total volumes of the samples are not equal. If we were to
keep the effective imaginary refractive indices constant, we would find
a significantly smaller absorption contribution in the nanopore film,
because it has a significantly smaller sample thickness. In this case,
keeping the volume constant in the two samples allows us to compare them
more easily. When the absorber is in the photonic-ball-film matrix, we
use the equation
\begin{equation} 
V_{\textnormal{cb}} = \frac{n_{\textnormal{i, PB-film matrix}}(1-\phi_{\textnormal{ball}})}{n_{\textnormal{i,cb}}}tA,
\end{equation}
where $V_{\textnormal{cb}}$ is the volume of carbon black in the sample,
$n_{\textnormal{i, PB-film matrix}}$ is the imaginary refractive index
of the photonic-ball-film matrix, $\phi_{\textnormal{ball}}$ is the
volume fraction of photonic balls in the film, the imaginary index of
the carbon black is
$n_{\textnormal{i,cb}}=0.44$~\cite{dalmeida_atmospheric_1991,
  chylek_effect_1995}, $t$ is the sample thickness, and $A$ is an
arbitrary sample area. When the absorber is placed in the photonic-ball
matrix, we use the equation
\begin{equation} 
V_{\textnormal{cb}}  = \frac{n_{\textnormal{i, PB matrix}}(1-\phi)}{n_{\textnormal{i,cb}}}tA,
\end{equation}
where $n_{\textnormal{i, PB matrix}}$ is the imaginary refractive index
of the photonic-ball matrix and $\phi$ is the volume fraction of
nanopores or nanoparticles. When the absorber is placed in the nanopores
or nanoparticles, we use the equation
\begin{equation} 
V_{\textnormal{cb}}  = \frac{n_{\textnormal{i, np}}\phi}{n_{\textnormal{i,cb}}}tA,
\end{equation} 
where $n_{\textnormal{i, np}}$ is the imaginary refractive index of the
nanopores or nanoparticles.

\subsection{Effective imaginary indices}

For the simulations shown in Fig.~7 of the main text, we keep the
effective, sample imaginary refractive indices constant across samples
with the broadband absorber placed in the three locations: the
photonic-ball-film matrix, the photonic-ball matrix, and the nanopores.
For this set of simulations, keeping the imaginary refractive indices
constant (rather than the total volume of carbon black) allows a more
fair comparison since the samples each have the same total volume.
However, we must be careful when comparing refractive indices across
samples where the absorber is placed in these different locations. When
the absorber is placed inside the photonic ball, either in the
photonic-ball matrix or the nanopores, we use the Bruggeman weighted
average to calculate an effective index for the photonic balls, which
takes into account the nanopore volume fraction and the complex
refractive indices of the photonic-ball matrix and
nanopores~\cite{markel_introduction_2016}. We take the imaginary
component of this index and then multiply by the volume fraction of
photonic balls in the film to approximate a refractive index for the
entire sample:
\begin{equation} 
n_{\textnormal{i, sample}} = \textnormal{Im}(n_{\textnormal{bruggeman}})\phi_{\textnormal{ball}},
\end{equation}
where $n_{\textnormal{bruggeman}}$ is the refractive index calculated
using the Bruggeman weighted average, and $\phi_{\textnormal{ball}}$ is
the volume fraction of photonic balls in the film.

When the absorber is in the photonic-ball-film matrix, we specify an
imaginary component of the refractive index and then multiply by the
total volume fraction of matrix material surrounding the photonic balls:
\begin{equation} 
n_{\textnormal{i, sample}} = n_{i, \textnormal{PB-film matrix}}(1-\phi_{\textnormal{ball}}),
\end{equation}
where $n_{i, \textnormal{PB-film matrix}}$ is the imaginary refractive
index of the photonic-ball-film matrix. Below, we report the imaginary
refractive indices used as inputs to the model for each of the
simulations shown in Fig.~7 of the main text, and we also list their
corresponding effective, sample imaginary indices, which we keep
constant across the simulations for different absorber locations:
\begin{center}
\begin{table}
\begin{tabular}{ cccc} 
 \toprule
 $n_{\textnormal{i, sample}}$ & $n_{\textnormal{i, PB-film matrix}}$ & $n_{\textnormal{i, PB matrix}}$ & $n_{\textnormal{i, nanopore}}$ \\ 
\midrule 
 \SI{8.25e-5}{} & \SI{1.65e-4}{} & \SI{3.83e-4}{} & \SI{2.86e-4}{} \\ 
 
 \SI{2.06e-4}{} & \SI{4.12e-4}{} &\SI{9.58e-4}{} & \SI{7.15e-4}{} \\ 

  \SI{3.30e-4}{} & \SI{6.59e-4}{} & \SI{1.53e-3}{} & \SI{1.14e-3}{} \\ 

 \SI{4.53e-4}{} & \SI{9.06e-4}{} &  \SI{2.11e-3}{} &  \SI{1.57e-3}{} \\ 

 \SI{5.77e-4}{} & \SI{1.15e-3}{} &  \SI{2.68e-3}{} &  \SI{2.00e-3}{}\\ 
 \bottomrule
 \caption{Imaginary index values for Fig. 7 of the main text}
 \end{tabular}
 \end{table}
 \end{center}

%%%%%%%%%%%%% 
\section{Additional results}
%%%%%%%%%%%%%

\begin{figure*} 
\centering \includegraphics{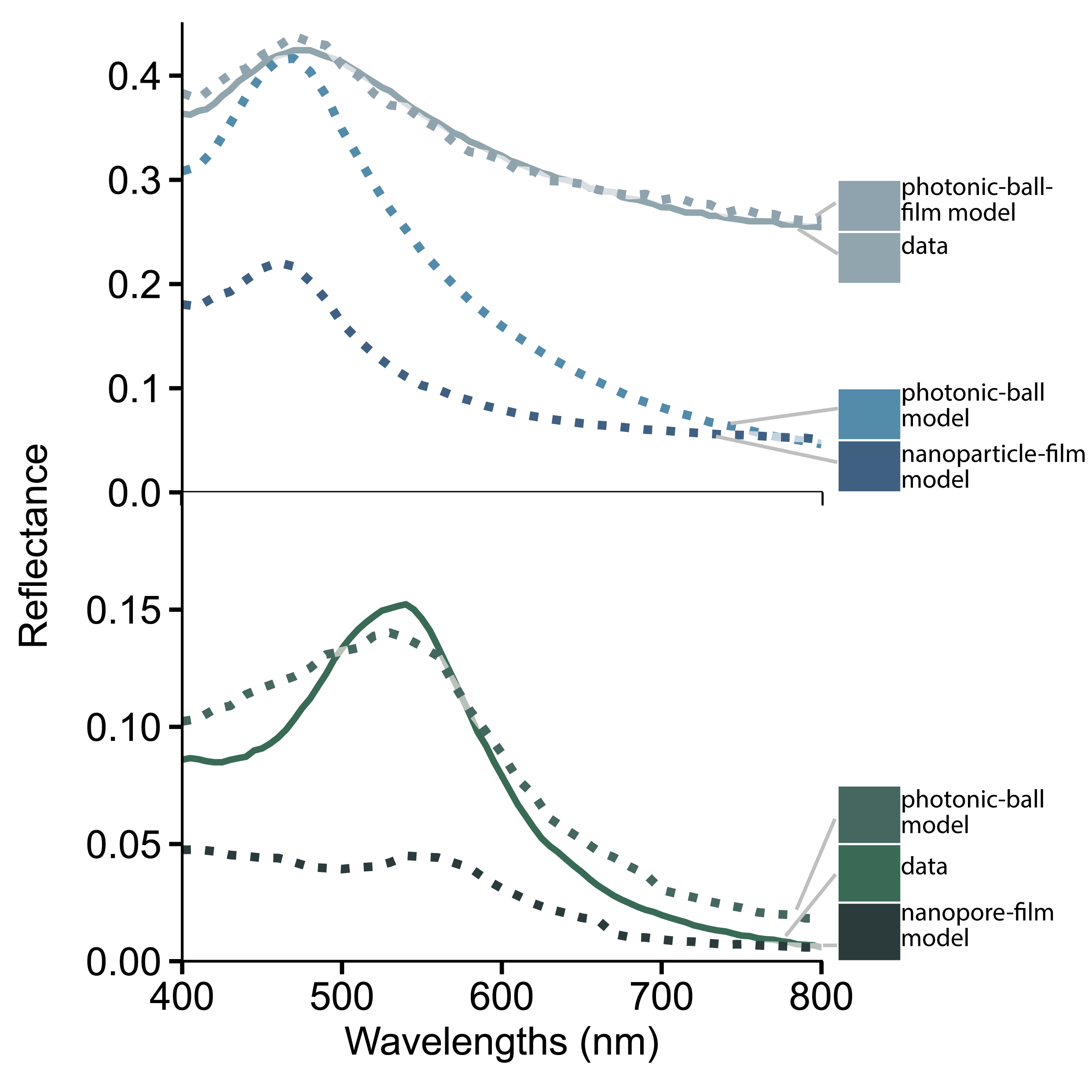}
\caption{Additional model comparisons for a photonic-ball film and a
  single photonic ball. \textit{Top}: Measured reflectance data (solid
  line) for a photonic-ball film and predicted reflectance spectra
  (dashed lines) for various geometries of our Monte Carlo models for a
  sample from Ref.~\citenum{takeoka_production_2013}. The measured
  reflectance and photonic-ball film model reflectance are also shown in
  Fig.~4b of the main text, and the model parameters are listed in the
  corresponding caption. For the nanoparticle film, the film thickness
  is multiplied by the volume fraction of photonic balls so that the
  volume of scattering material is the same as in the photonic-ball film
  models: thickness $= \SI{200}{\micro\meter} \times 0.55 =
  \SI{110}{\micro\meter}$. The imaginary refractive index for the matrix
  of the nanoparticle film is \SI{3.997e-4}{}. This imaginary index is
  chosen to keep the volume the same across the models for the
  nanoparticle film and photonic-ball film. Other parameters are shared
  between the simulations and are listed in the caption of Fig.~4b of
  the main text. \textit{Bottom}: Measured reflectance data (solid line)
  for an individual photonic ball and predicted reflectance spectra
  (dashed lines) for two geometries of Monte Carlo models for an
  individual photonic ball sample from
  Ref.~\citenum{zhao_angular-independent_2020}. The measured reflectance
  and photonic-ball film model reflectance are also shown in Fig.~3b of
  the main text, and the model parameters are listed in the
  corresponding caption.}
\label{S1}
\end{figure*}

\begin{figure*} 
\centering \includegraphics{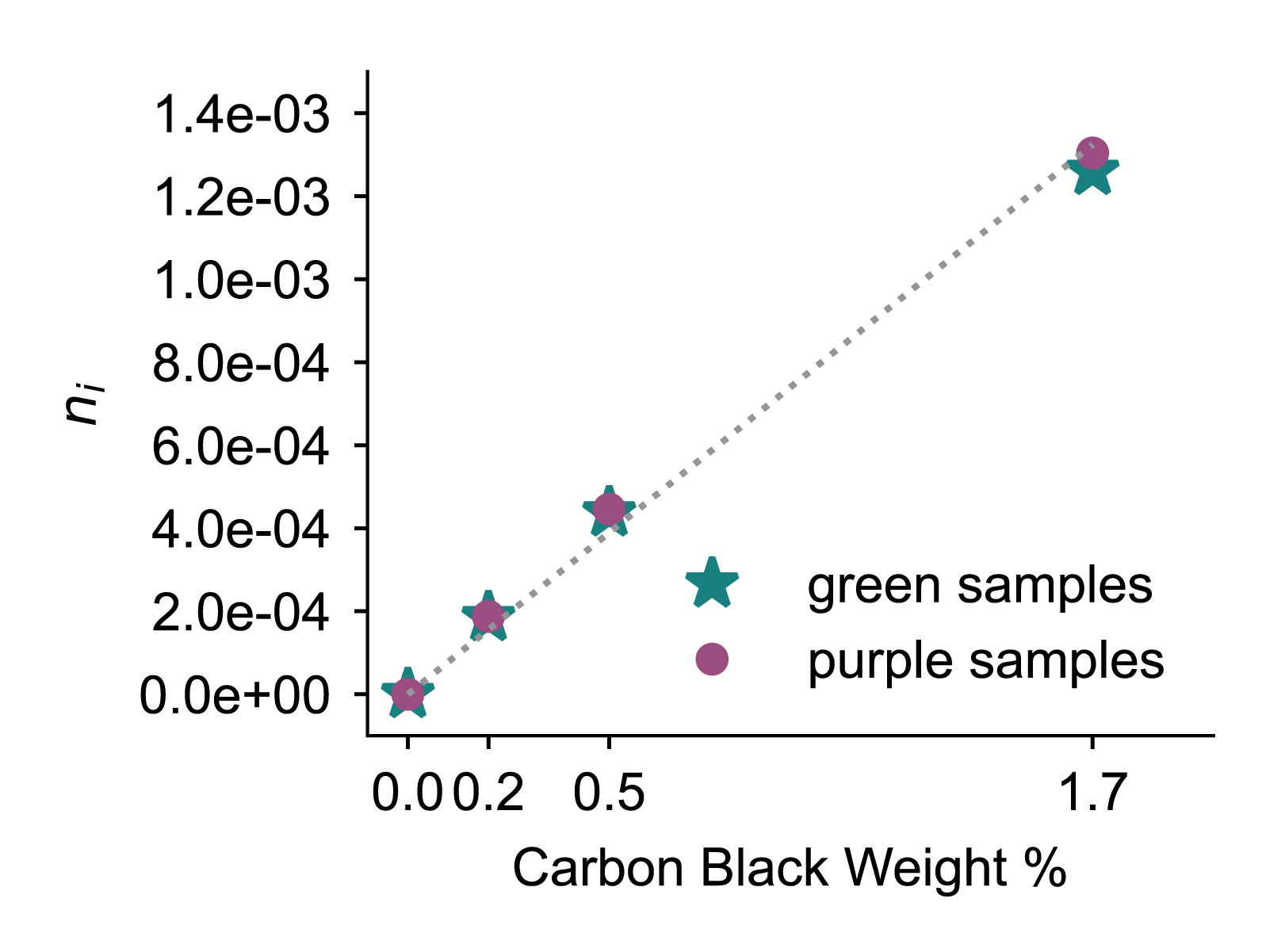}
\caption{Imaginary index increases linearly with carbon black
  concentration. Plot shows the best-fit imaginary index for the green
  samples (stars) and purple samples (circles) for the samples from
  Ref.~\citenum{takeoka_production_2013}, with reflectance spectra
  plotted against our model in Fig.~4b of the main text. The dashed line
  shows a least-squares fit for the purple samples, with equation
  ${n_i}$ = \SI{7.79e-4} {[cb]}, where $n_i$ is the imaginary index and
  {[cb]} is the concentration of carbon black. The plotted $n_i$ values
  combine the carbon black concentrations in the photonic-ball matrix
  with those in the photonic-ball-film matrix by weighting them
  according to volume fraction and adding the two values. For the green
  samples, the $n_i$ values for the photonic-ball matrix are, from
  darkest to lightest color: \SI{1e-6}, \SI{4e-6}, \SI{3e-6}, 0, and the
  $n_i$ values for the photonic-ball-film matrix are \SI{2.8e-3},
  \SI{9.7e-4}, \SI{4.1e-4}, 0. For the purple samples, the $n_i$ values
  for the photonic-ball matrix are, from darkest to lightest color:
  \SI{1.2e-4}, \SI{3.7e-5}, \SI{3e-6}, 0, and the $n_i$ values for the
  photonic-ball-film matrix are \SI{2.83e-3}, \SI{9.7e-4},
  \SI{4.15e-4}, 0.}
\label{S2}
\end{figure*}

Fig.~\ref{S1} shows comparison between measurements and various model
predictions for additional samples, and Fig.~\ref{S2} shows the linear
relationship between the sample's carbon black concentration and the
sample's imaginary index for the imaginary indices input into
simulations shown in Fig.~4b of the main
text.

%%%%%%%%%%%%%%%%%%%%%%%%%%%%% 
\section{Investigating the effective-medium approximation} 
%%%%%%%%%%%%%%%%%%%%%%%%%%%%%

The use of effective-medium theory for materials with refractive indices
on the order of the wavelength of light is an active area of research
and deserves some discussion as it relates to our model. For low
index-contrast samples, such as polystyrene particles in water,
effective-medium theory is normally not needed to accurately predict
scattering properties~\cite{fraden_multiple_1990,
  kaplan_diffuse-transmission_1994}. However, as index contrast
increases, such as in samples of polymers in air, an effective-medium
approximation can improve predictions of the scattering resonances. To
illustrate how the Bruggeman effective-medium theory, which we use in
our model, affects the prediction of scattering resonances and
scattering strength, we compare the scattering strength predicted by our
model to data on scattering strength as a function of size parameter for
polystyrene particles in air. The scattering strength is defined as
$1/l*$, where $l*$ is the transport length. These data are from Fig.~4
of Ref.~\citenum{aubry_resonant_2017} and from
Ref.~\citenum{garcia_resonant_2008}.

\begin{figure*} 
\centering \includegraphics{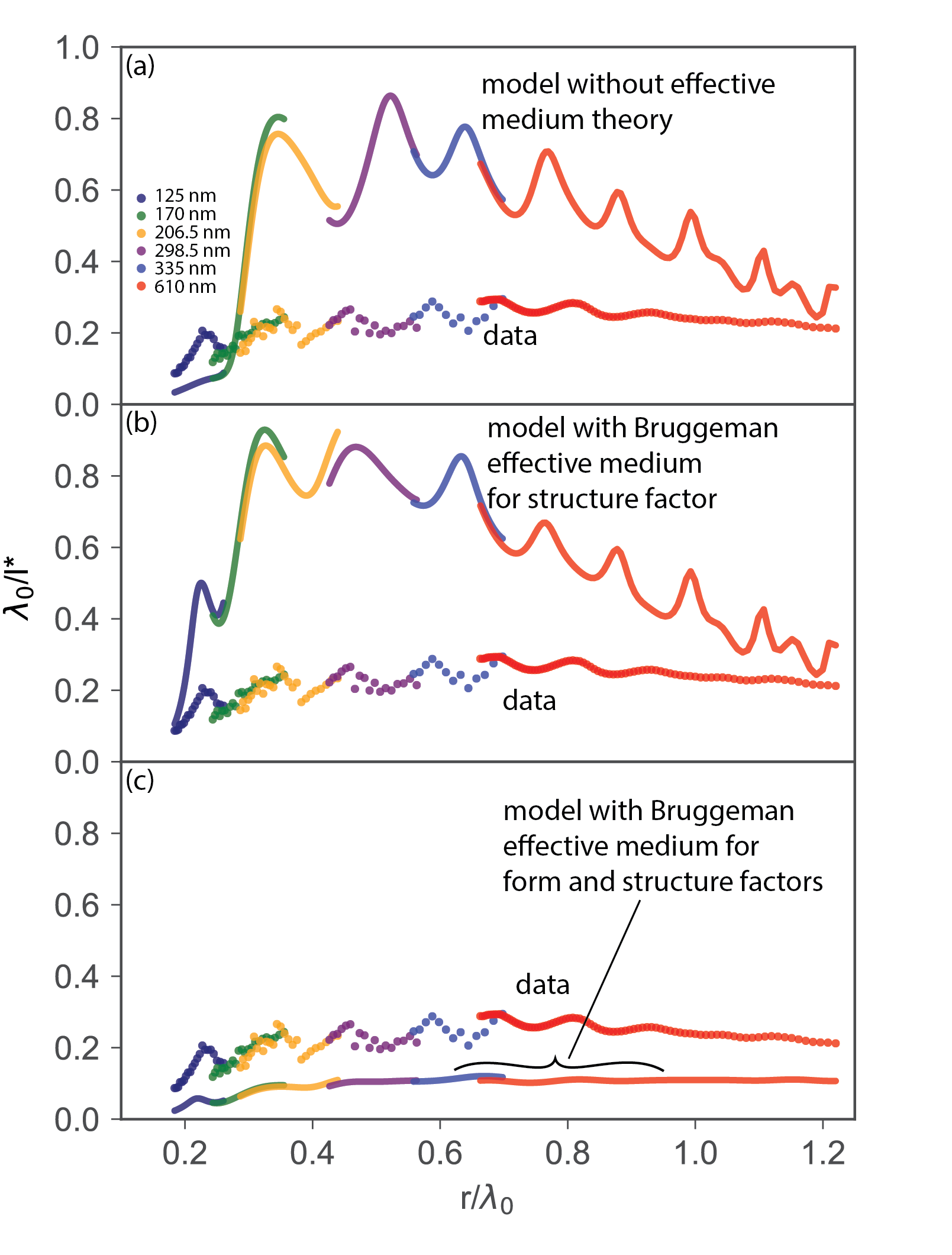}
\caption{Measured and calculated scattering strength of polystyrene
  particles with radius \SIrange{125}{610}{\nano\meter} and a volume
  fraction of 0.5. Calculations use refractive indices corresponding to
  (a) polystyrene particles in air with no effective-medium
  approximation, (b) polystyrene particles in a Bruggeman effective
  medium applied to the structure-factor calculation but in air for the
  form-factor calculation, and (c) polystyrene particles in a Bruggeman
  effective medium applied to the structure-factor and form-factor
  calculations. Data is reproduced from
  Ref.~\citenum{garcia_resonant_2008} (red circles) and
  Ref.~\citenum{aubry_resonant_2017} (other colors).}
\label{S3}
\end{figure*}

Without effective-medium theory, the calculation overestimates the
scattering strength by up to a factor of 4 (Fig.~\ref{S3}a); also, the
peaks in the scattering strength, corresponding to resonances, do not
align with those in the data. Next, we use the Bruggeman
effective-medium approximation in only the structure factor, which
allows us to account for the effect of the phase delay as described in
recent work by Yazhgur and colleagues~\cite{yazhgur_scattering_2022}.
This approach helps to align the resonances with those of the data, but
the magnitudes of the scattering strength are still overestimated
(Fig.~\ref{S3}b). Finally, we show the results of including the
effective-medium in both the form and structure factor, as is the case
in our model (Fig.~\ref{S3}c). Although the effective-medium approach
underestimates the scattering strength, it aligns the resonances with
the data for the size parameters we are interested in (roughly less than
0.5) and also gives a closer magnitude for the scattering strength.
These calculations show that the use of the Bruggeman effective index
for the form and structure factor matches more closely to data than the
other options for samples with an index contrast corresponding to
polystyrene in air. Moreover, we note that many researchers have
successfully used effective-medium theories to predict scattering
properties of materials in this intermediate refractive index
regime~\cite{prum_coherent_1998, forster_biomimetic_2010,
  hwang_designing_2021}.

\begin{figure*} 
\centering \includegraphics{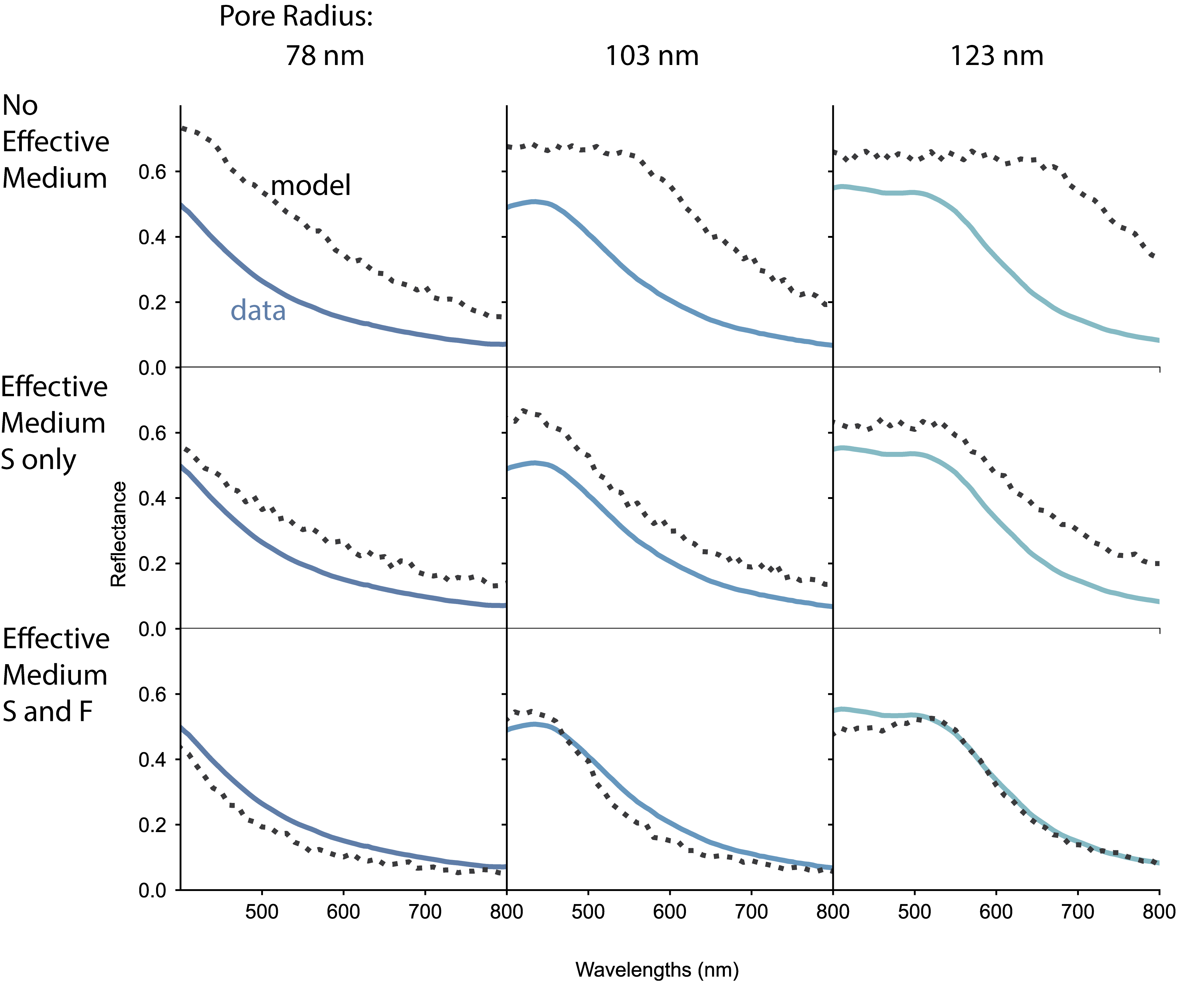}
\caption{ Measured (solid lines) and predicted (dotted lines)
  reflectance spectra for photonic balls with a primary reflectance peak
  in the ultraviolet, blue, and green. Measurements are from
  Ref.~\citenum{zhao_angular-independent_2020}. Each row compares
  measurements to model predictions with a different sample refractive
  index. The model calculations for each row are \textit{top}, no
  effective index; \textit{middle}, Bruggeman effective refractive index
  used to calculate the structure factor only; \textit{bottom},
  Bruggeman effective refractive index used to calculate both the
  structure and form factors. The model parameters are \textit{left}:
  nanopore radius \SI{78}{\nm}, photonic ball diameter \SI{40}{\um},
  \textit{middle}: nanopore radius \SI{103}{\nm}, photonic ball diameter
  \SI{19.9}{\um}, and \textit{right}: nanopore radius \SI{123}{\nm},
  photonic ball diameter \SI{19.3}{\um}. All simulations use a nanopore
  volume fraction of 0.5, a fine roughness of 0.01, a matrix refractive
  index of 1.52, and a matrix and medium refractive index of 1.}
\label{S4}
\end{figure*}

To further compare these three different types of calculations, we ran
the model for the samples from Zhao and
colleagues~\cite{zhao_angular-independent_2020} shown in Fig.~3 in the
text. These samples have an index contrast of roughly $0.52$. We find
that the peak position is predicted accurately when an effective-medium
approximation is used for the structure factor only and when an
effective-medium approximation is used for both the form factor and
structure factor (Fig.~\ref{S4}). For these measurements we choose a
measurement aperture half-angle of \ang{60} to fit the magnitude of the
data. Because this number is a fitting parameter, it is difficult to
determine whether the second approach (effective medium for structure
factor only) or third approach (effective medium for both form and
structure factor) is a better approximation. However, based on the
scattering strength calculations above, using the full Bruggeman
approximation appears to be the best choice for an intermediate index
contrast since it produces a scattering strength closer to
the data.

\begin{figure*} 
\centering \includegraphics{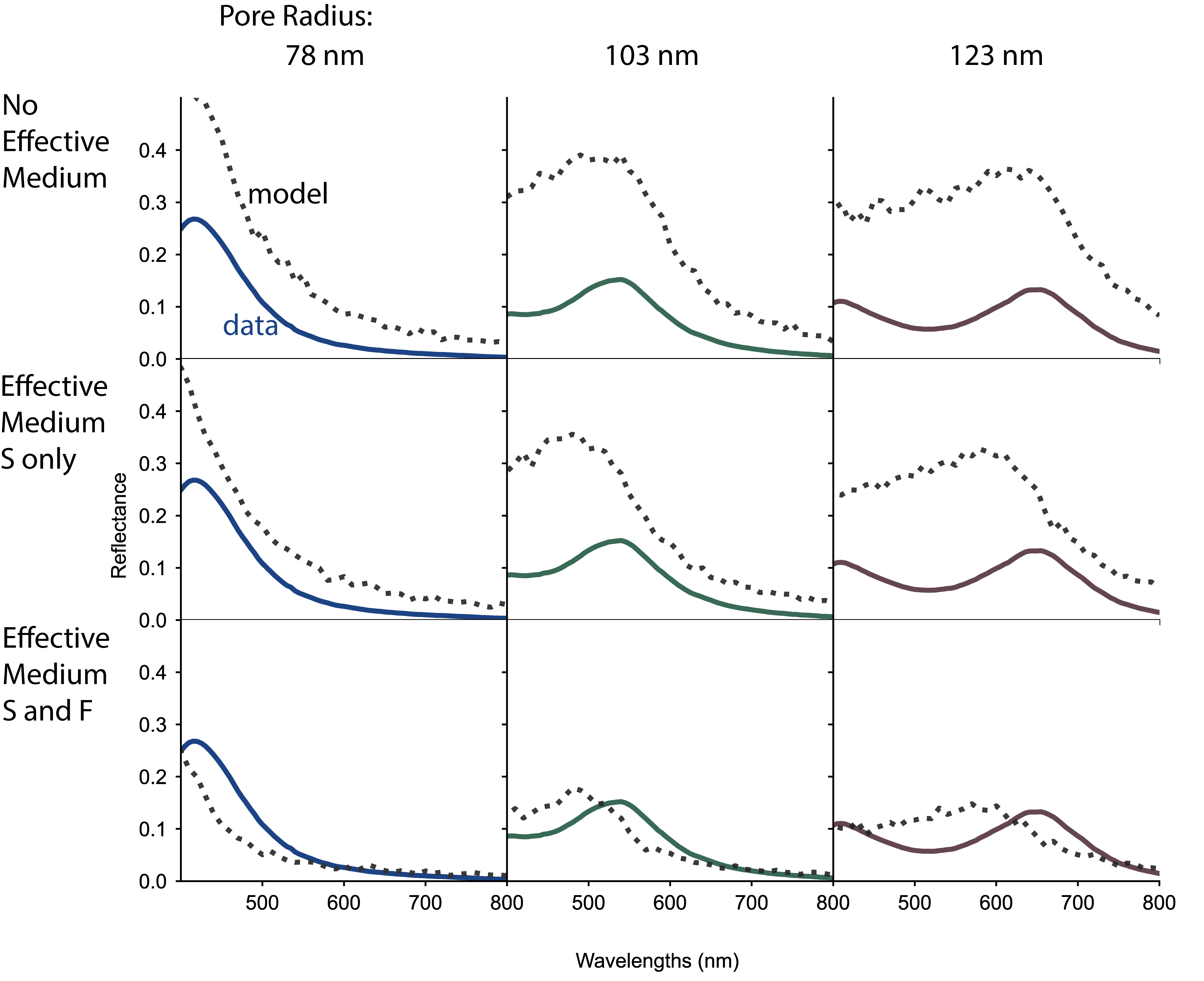}
\caption{Measured (solid lines) and predicted (dotted lines) reflectance
  spectra for photonic balls with a primary reflectance peak in the
  blue, green, and red. Measurements are from
  Ref.~\citenum{zhao_angular-independent_2020}, and calculations are as
  described in Fig.~\ref{S4}. The model parameters are \textit{left}:
  nanopore radius \SI{78}{\nm}, photonic ball diameter \SI{40}{\um},
  \textit{middle}: nanopore radius \SI{103}{\nm}, photonic ball diameter
  \SI{19.9}{\um}, and \textit{right}: nanopore radius \SI{123}{\nm},
  photonic ball diameter \SI{19.3}{\um}. All simulations use a nanopore
  volume fraction of 0.5, a fine roughness of 0.01, a matrix refractive
  index of 1.52, and a matrix and medium refractive index corresponding
  to that of water.}
\label{S5}
\end{figure*}

For completeness, we also use the same three types of calculations to
model the reflectance of the same photonic-ball samples (Fig.~\ref{S5})
immersed in an aqueous suspension, which lowers their index contrast to
roughly $0.19$. As expected, we find agreement between the measured and
predicted peak locations only when we do not use effective-medium
theory.

\clearpage
\bibliography{references}